\documentclass[prl,showpacs,twocolumn]{revtex4}
\usepackage{graphicx}
\usepackage{dcolumn}
\usepackage{bm}
\usepackage{amssymb}
\usepackage{amsmath}
\usepackage{amsfonts}
\usepackage{soul}

\usepackage[]{graphicx}
\def\beq{\begin{eqnarray}}
\def\eeq{\end{eqnarray}}

\begin{document}

\title{Imaging surface plasmons: from leaky waves to far-field radiation}
\author{Aur\'elien Drezet}
\affiliation{Institut N\'eel, UPR 2940, CNRS-Universit\'e Joseph
Fourier, 25, rue des Martyrs, 38000 Grenoble, France}
\author{Cyriaque Genet}
\affiliation{ISIS, UMR 7006, CNRS-Universit\'e de Strasbourg, 8,
all\'ee Monge, 67000 Strasbourg, France}
\date{\today}

\begin{abstract}
We show that, contrary to the common wisdom, surface plasmon poles
are not involved in the imaging process in leakage radiation
microscopy. Identifying the leakage radiation modes directly from a
transverse magnetic potential leads us to reconsider the surface
plasmon field and unfold the non-plasmonic contribution to the image
formation. While both contributions interfere in the imaging
process, our analysis reveals that the reassessed plasmonic field
embodies a pole mathematically similar to the usual surface plasmon
pole. This removes a long-standing ambiguity associated with
plasmonic signals in leakage radiation microscopy.

\end{abstract}
\pacs{42.25.Lc, 42.70.-a, 73.20.Mf} \maketitle

\indent Surface plasmon optics has become a mature and extended
field of research, ranging from the development of new optical
nanodevices and nanoantennas to the renewal of integrated quantum
optics \cite{BarnesNature2003,AgioNano2012}. In this context,
surface plasmon imaging techniques are of critical importance to the
researcher, among which leakage radiation microscopy (LRM) is now
emerging as a powerful tool~\cite{Hecht,DrezetMatResB2008}. As a
far-field optical method, LRM is used for analyzing SP modes both in
direct and Fourier (momentum) spaces and it has been successfully
implemented in various plasmonic systems, both at the classical and
quantum level
\cite{BaudrionOptX2008,GorodetskiPRL2012,LiPRL2013,BharadwajPRL2011,CucheNL2010,MolletPRB2012}.
Yet, there is still no satisfying theoretical definition of the SP
field in an imaging context. Instead, recent reference work on leaky
waves have focused on semi-infinite air-metal interfaces, a
configuration not relevant  to LRM
\cite{CollinIEEE2004,NikitinNJP2009,LalanneSSR2009,NikitinPRL2010}.
This has fuelled recent debates concerning the precise relation
between experimentally recorded images and
SP modes \cite{WangOL2010,DePeraltaOL2010,HohenauOptX2011}. \\
\indent In this Letter, we instead propose a novel approach to the
problem of leaky waves that provides a full analytical theory of the
coherent SP imaging process in the case of a point-like radiating
electric dipole located in air above the thin metal film. This leads
us to  a new definition of the SP field as a Fano-type interfering
component of the imaged radiation. We derive analytical expressions
for the far-field radiation that meets all the necessary conditions
prescribed to the leakage field symmetries \cite{BurkePRB1986}.
Importantly, we show that our approach naturally makes the SP field
free from the long standing ambiguities of the historical Zenneck
and Sommerfeld solutions \cite{ZenneckAP1907,SommerfeldAP1909} and
removes the field discontinuity at the LR angle, until now
problematic. Doing so, we also identify the contribution of a
lateral wave thus far unnoticed that we associate with a new type of
Goos-H\"anchen effect in transmission.\\
\begin{figure}[htb]
\centering
\includegraphics[width=8cm]{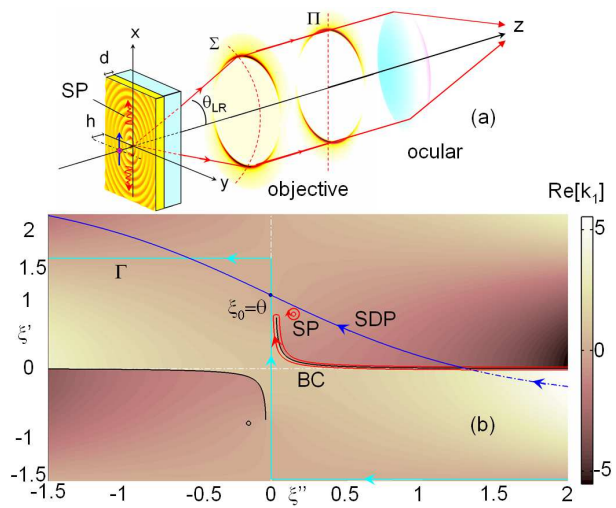}
\caption{(a) Sketch of the leakage radiation microscope: an electric
dipole embedded in medium 1 with permittivity $\varepsilon_1= 1$
(air) is located at $[x=0,y=0,z=-h]$ from the surface $z=0$ of a
thin (thickness $d$) metal film of permittivity $\varepsilon_2$
deposited on a glass substrate of permittivity $\varepsilon_3=n^2
>\varepsilon_1$ (see \cite{footLeg1}). (b) Integration path
$\Gamma$ in the $(\xi''={\rm Im}(\xi),\xi'={\rm Re}(\xi))$ complex
plane. BC is the branch-cut  and SP the position of the plasmonic
poles (symmetric leaky) in the two ${\rm Im}(\xi)\cdot {\rm
Re}(\xi)>0$ quadrants. The steepest descent path SDP is shown in the
${\rm Im}(k_1)>0$ first $R_+$ Riemann sheet (continuous line), and
in the ${\rm Im}(k_1)<0$ second Riemann sheet (dot-dashed). The
variations of ${\rm Re}(k_1)$ are displayed with values given in the
color-bar.}  \label{fig1}
\end{figure}
\indent In the geometry considered in Fig.~\ref{fig1}, a harmonically
radiating point-like dipole $\boldsymbol{\mu}e^{-i\omega t}$ drives an SP wave that leaks through the
film in medium 3 and then propagates in the matching oil of the
high numerical aperture (NA) immersion objective required for LRM. Due
to the specific dispersion relation of SP waves, leakage radiation (LR) is
emitted at an angle
$\vartheta_{\textrm{LR}}>\vartheta_c=\arcsin{[\sqrt{(\varepsilon_1/\varepsilon_3)}]}$
defining a radiation cone  in the forbidden-light sector shown on
Fig.~\ref{fig1}(a) which intersects the reference sphere $\Sigma$ of the LRM objective.\\
\indent Due to the planar
symmetry of the problem, the radiated field in medium 3 can be represented in terms of transverse magnetic
(TM) and electric (TE) scalar potentials $\Psi$ with
$[\boldsymbol{\nabla}^2+k_0^2\varepsilon_3]\Psi_{\textrm{TM,TE}}=0$ where $k_0=\omega/c$.
For simplicity, the case of a dipole normal to the film
$\boldsymbol{\mu}=\mu_\bot\hat{\mathbf{z}}$ (i.e.
$\Psi_{\textrm{TE}}=0$) is only discussed here. A general and detailed calculation is given in~\cite{supll}. Using boundary conditions
at the different interfaces, we expand the potential at $[\mathbf{x}=(x,y),z]$ as
$\Psi_{\textrm{TM}}(\mathbf{x},z)=\int_{-\infty}^{+\infty} kdk
A(k,z)H_0^{(+)}(k|\mathbf{x}|)$ with $A(k,z)=\frac{i\mu_\bot}{8\pi
k_1}\tilde{T}_{13}^{\textrm{TM}}(k)e^{ik_1h}e^{ik_3z}$ and
$H_0^{(+)}(k|\mathbf{x}|)$ is the zeroth order radiating-like Hankel
function (evolving asymptotically as
$e^{+ik|\mathbf{x}|}/\sqrt{k|\mathbf{x}|}$ for large $|\mathbf{x}|$). The Fresnel coefficient
$\tilde{T}_{13}^{\textrm{TM}}(k)$ gives the transmission through the film of TM radiation with a wave vector
$|\mathbf{k}|=k$. Stability imposes complex square roots $k_j(k)=\sqrt{k_0^2\varepsilon_j-k^2}$
($j=1$ or 3) with $\textrm{Im}[k_j]\geq0$. \\
\indent The precise computation of such a Sommerfeld-like integral
is extraordinarily involved due to the presence of two branch-cuts
associated with $k_{1,3}(k)$ and several SP poles in the complex
$k$-plane \cite{note1}. To simplify at most the problem, we chose an
alternative parametrization of the integral through the complex
variable $\xi$ defining $k=k_0n\sin{\xi}$. This leaves only one
branch-cut $k_1=k_0\sqrt{\varepsilon_1-\varepsilon_3\sin^2{\xi}}$
with a branch-point $\xi= \vartheta_c$. We impose
$\textrm{Im}[k_1]\geq0$ in the whole complex $\xi$-plane as the
choice of the Riemann sheet $R_+$. The integral becomes
\begin{eqnarray}
\Psi_{\textrm{TM}}(\mathbf{x},z)=\int_{\Gamma}d\xi
F(\xi)e^{ik_0nr\cos{(\xi-\vartheta)}}\label{field}
\end{eqnarray}
where
$F(\xi)=A(k,d)kk_3H_0^{(+)}(k|\mathbf{x}|)e^{-ik|\mathbf{x}|}$ with the polar
coordinates $|\mathbf{x}|=r\sin{\vartheta}$, $z=d+r\cos{\vartheta}$
leading to
$(z-d)\cos{\xi}+|\mathbf{x}|\sin{\xi}=r\cos{(\xi-\vartheta)}$. The
initial contour $\Gamma$ corresponding to the condition $\sin{\xi}$ real is
represented on Fig.~1(b).   \\
\indent To evaluate Eq.~(\ref{field}), we deform the contour $\Gamma$ in order to
include the steepest descent path (SDP) determined by
$\textrm{Im}[f_\vartheta(\xi)]=1$ with
$f_\vartheta(\xi)=i\cos(\xi-\vartheta)$. The SDP crosses $\Gamma$ at the
saddle point $\xi_0=\vartheta$ defined by
$df(\xi)/d\xi=0$. The SDP contribution $\Psi_{\textrm{SDP}}$ to the field is calculated
using the steepest-descent method discussed below. Two additional contributions $\Psi_p$ and
$\Psi_{\textrm{BC}}$ associated respectively with the SP poles and the BC have to be accounted for when deforming the contour. The most relevant part here is a single SP pole
$k_p=k_0n\sin{\xi_p}$ resulting from the divergency of
$\tilde{T}_{13}^{\textrm{TM}}(k)=\frac{N_{13}(k)}{D_{13}(k)}$ when $D_{13}(k_p)=0$.
Such a transcendental equation is known to
possess four kinds of SP modes corresponding to leaky waves and bound
modes in medium $1$ or $3$~\cite{BurkePRB1986,DrezetMatResB2008}. Importantly on the $R_+$ sheet,
only the leaky mode in medium $3$ (labeled symmetric
leaky in~\cite{BurkePRB1986}) is possibly encircled
during the contour integration (see Fig.~\ref{fig1}(b)) depending on whether
$\xi_0>\vartheta_{\textrm{LR}}$ or not. This implies that the residue contribution to $\Psi_{\rm TM}$ associated
with the SP pole reads
\begin{eqnarray}
\Psi_{p}=2\pi i
\textrm{Res}[F(\xi_p)]e^{ik_0nr\cos{(\xi_p-\vartheta)}}\Theta(\vartheta-\vartheta_{\textrm{LR}}),
\end{eqnarray}
where $\Theta(x)$ is the Heaviside unit-step function and where the LR angle is precisely defined by $\textrm{Im}[f_{\vartheta_{\textrm{LR}}}(\xi_p)]=1$, i.e. $\vartheta_{\textrm{LR}}=\textrm{Re}[\xi_p]+\arccos{(1/\cosh{\textrm{Im}[\xi_p]})}\simeq \textrm{Re}[\xi_p]$.
\begin{figure}[htb]
\centering
\includegraphics[width=6cm]{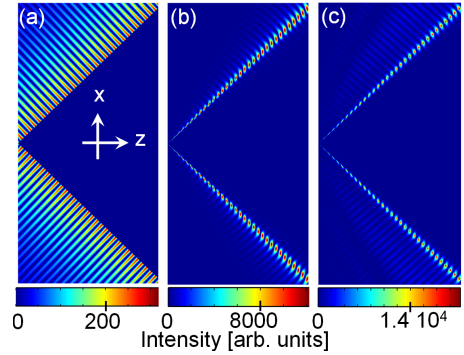}
\caption{Intensity contour-plots corresponding to the contributions
(a) $\Psi_p$ and (b) $\Psi_{\textrm{SDP}}$, with their coherent
superposition displayed in (c). To remove the asymptotic divergence
at infinity we calculated $|\mathbf{x}|(\textrm{Re}[\mathbf{E}])^2$
in (a), while  we calculated instead
$r^2(\textrm{Re}[\mathbf{E}])^2$ in (b, c).} \label{fig2}
\end{figure}
An intensity plot of this contribution is displayed on Fig.  \ref{fig2}(a) which clearly shows the conical wave front
structure emitted at the angle
$\arctan{(\textrm{Re}[k_p]/\textrm{Re}[k_{3p}])}=\textrm{Re}[\xi_p]\simeq\vartheta_{\textrm{LR}}$. Remarkably, this conical
$\Theta(\vartheta-\vartheta_{\textrm{LR}})$ wave front removes the power flow divergence in the $z>0$ direction due to the exponentially growing field associated with $\Psi_p$ in medium $3$~\cite{BurkePRB1986} and the spatial singularity along the $z$ axis
surviving at all distances away from the source induced by the Hankel function~\cite{supll}. \\
\indent Beyond
$\vartheta_{\textrm{LR}}$,
$\Psi_p\propto e^{ik_{3p}z}H_0^{(+)}(k_p|\mathbf{x}|)$ (where
$k_{3p}:=k_3(k_p)$) in direct relation to the original
derivation by Zenneck and Sommerfeld of surface
waves~\cite{ZenneckAP1907,SommerfeldAP1909}. It corresponds to what
modern literature coined leaky SP
mode~\cite{BurkePRB1986,DrezetMatResB2008}, belonging to the
general family of leaky waves discussed for years in the radio
antenna community~\cite{JacksonBook2008}. It is important to realize however that this contribution is actually non-physical, due to the field discontinuity at $\vartheta_{\textrm{LR}}$ introduced by the $\Theta(\vartheta-\vartheta_{\textrm{LR}})$ pre-factor. It thus appears necessary to find a genuinely physical definition of a leaky SP wave. We now show that this is only possible by including in our discussion both $\Psi_{\textrm{SDP}}$ and $\Psi_{\textrm{BC}}$ contributions.  \\
\indent The central result of the Letter is that the LRM imaging process is essentially determined by $\Psi_{\textrm{SDP}}$ which is evaluated as~\cite{supll}
\begin{eqnarray}
\Psi_{\textrm{SDP}}=e^{ik_0nr}\sum_{m\in
\textrm{even}}\frac{\Gamma(\frac{m+1}{2})}{m!(k_0nr)^{\frac{m+1}{2}}}\frac{d^m}{d\tau^m}G(0)\label{taylor}
\end{eqnarray}
with the variable $\tau=e^{i\pi/4}\sqrt{2}\sin{((\xi-\vartheta)/2)}$
and the function $G(\tau)=F(\xi)\frac{d\xi}{d\tau}$ in the vicinity
of the saddle point $\tau=0$ (i.e. $\xi_0=\vartheta$). A second
contribution must be accounted for when $\vartheta>\vartheta_c$
because in this case, the close integration path have to surround
the branch-cut in the $R_+$ sheet (see Fig.~\ref{fig1}(b)).
Following~\cite{BrekhovskikhBook}, this contribution can be given as
a series expansion
\begin{eqnarray}
\Psi_{\textrm{BC}}\approx
e^{i\Delta\varphi}\Theta(\vartheta-\vartheta_c)\sum_{m=1}^{+\infty}\frac{\alpha_m(r,\vartheta_c)}{(k_0nr\sin{(\vartheta-\vartheta_c)})^{1+m/2}}  \label{eqBC}
\end{eqnarray}
where the coefficients $\alpha_m(r,\vartheta_c)$ can be explicitly
computed and $\Delta\varphi=k_0nr\cos{(\vartheta-\vartheta_c)}$ is
interpreted as the phase accumulated by a wave creeping along the
metal film (at velocity $c$) and re-emitted at the critical angle
$\vartheta_c$ (at velocity $c/n$)~\cite{supll}. Such a wave is
associated in our case to a
Goos-H\"{a}nchen-like effect in transmission~\cite{BrekhovskikhBook}. This contribution is not specific to the LRM geometry  but it has never been discussed  whereas the $m=1$ far-field dominant term in Eq.~(\ref{eqBC}) evolves as $\sim 1/r^2$, i.e. as a Norton wave defined on the same $1/r^2$ order from $\Psi_{\textrm{SDP}}$~\cite{Norton}.  \\
\indent These terms however can be neglected in the far field ($r\gg 2\pi/k_0$) where only survives the dominant $1/r$ term in the power expansion of $\Psi_{\textrm{SDP},m=0}$ in Eq.~(\ref{taylor}). The radiated far field is thus
\begin{eqnarray}
\Psi_{\textrm{SDP},m=0}(\mathbf{x},z)\simeq\frac{2\pi
k_0n\cos{\vartheta}}{ir}e^{ik_0
nr}\tilde{\Psi}_{\textrm{TM}}[\mathbf{k},d]  \label{SDPm0}
\end{eqnarray}
where $\tilde{\Psi}_{\textrm{TM}}[\mathbf{k},d]=A(k,d)/\pi$ is the
bidimensional Fourier transform of
$\Psi_{\textrm{TM}}(\mathbf{x},z)$ calculated at $z=d$ for the
in-plane wave vector
$\mathbf{k}=k_0n\sin{\vartheta}\mathbf{x}/|\mathbf{x}|$. This
expression shows in Fig.~\ref{fig2}(b) a conical structure peaked on
$\vartheta_{\textrm{LR}}$ which should be compared with the one
obtained from $\Psi_p$ alone in Fig.~\ref{fig2}(a). The comparison
with Fig.~\ref{fig2}(c) combining both contributions coherently
clearly expresses a central result for LRM:
$\Psi_{\textrm{SDP},m=0}$ strongly dominates not only over
$\Psi_{\textrm{BC}}$, as discussed above, but also over $\Psi_p$,
consistently with the finite value of the SP propagation length
$L_p=(2\textrm{Im}[k_p])^{-1}$ that overdamps the exponential tail
of
$\Psi_p$ for angles $\vartheta\geq\vartheta_{\textrm{LR}}$.  \\
\begin{figure}[htb]
\centering
\includegraphics[width=8.6cm]{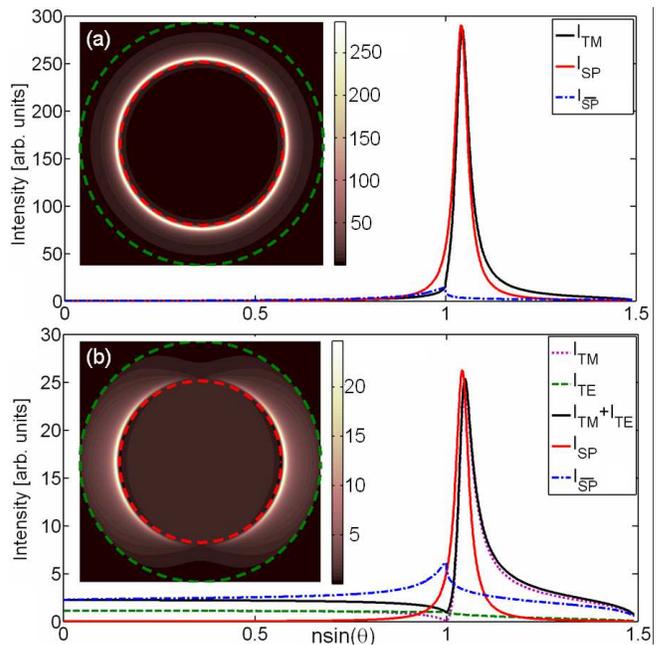}
\caption{Fourier space images recorded in the back-focal plane of
the microscope objective for a dipole radiating through a thin metal
film, when the dipole is perpendicular (a) or parallel (b) to the
metal film. In both cases, the calculated accessible intensities in
the $\mathbf{k}=[k_x,k_y]$-plane $I_{\textrm{TM,TE}}$ are associated with
$\tilde{\Psi}_{\textrm{TM,TE}}$ and $I_{\textrm{SP}}$ with the SP
contribution $\tilde{\Psi}_{\textrm{SP}}$. The non-plasmonic term
$I_{\overline{\textrm{SP}}}$ is defined as a residual $\propto
k_0^3k_3k^2\varepsilon_3|\tilde{\Psi}_{\textrm{TM}}-\tilde{\Psi}_{\textrm{SP}}|^2$
(a) and $\propto k_0^3k_3k^2\varepsilon_3|\tilde{\Psi}_{\textrm{TM}}-\tilde{\Psi}_{\textrm{SP}}|^2+I_{\textrm{TE}}$
(b). }  \label{fig3}
\end{figure}
\indent From an imaging perspective, the radiating field given by
Eq.~(\ref{SDPm0}) is detected in the objective back-focal plane
$\Pi$ sketched in Fig.~\ref{fig1} (see ~\cite{footLeg1}).  We plot
$I_{\textrm{TM},\Pi}[\mathbf{k}]$ in Fig.~\ref{fig3} for a dipole
either perpendicular or parallel to the interface~\cite{footLeg2}.
We point out that these intensity maps are in quantitative agreement
with
experiments~\cite{DrezetMatResB2008,BharadwajPRL2011,MolletPRB2012}
and clearly reveal a bidimensional ring with radius
$k_r\approx\textrm{Re}[k_p]$ and width $\delta k\approx
2\textrm{Im}[k_p]$. \\
\indent As a fundamental paradox, this ring is nowadays associated with the
detection of the SP mode~\cite{Novotny2,DrezetMatResB2008,Marty}
despite the fact that it is $\Psi_{\textrm{SDP}}$
and certainly not $\Psi_p$ which is involved in the measurement process. This paradox stems from $G(\tau)$ being singular (thus $\Psi_{\textrm{SDP}}$ too) at the SP pole $\tau_p$ (i.e. $\xi_p$), with a polar
contribution $\Psi_{\textrm{SDP}}^{pole}$ sharing a closed mathematical
relation with $\Psi_p$. This generated an historical
confusion regarding the actual role of SP modes in the Sommerfeld
integral, an issue debated since the work of Zenneck and
Sommerfeld~\cite{ZenneckAP1907,SommerfeldAP1909,CollinIEEE2004}.  \\
\indent In order to remove the ambiguity, we reconsider the very definition of what the SP field is, by going
back to Eq.~(\ref{taylor}) and observing that
$\tilde{\Psi}_{\textrm{TM}}[\mathbf{k},d]$ is an explicit function of
$k=k_0n\sin{\vartheta}$ with $\vartheta\in[0,\pi/2]$. This function can be
easily continued over the complex $k$-plane analytically. This function presents some isolated poles such as
$k_p$ and $-k_p$ which allow us to decompose
$\tilde{\Psi}_{\textrm{TM}}[\mathbf{k},d]$ into a regular and polar part. It is this
polar part
\begin{eqnarray}
\tilde{\Psi}_{\textrm{TM}}^{pole}[\mathbf{k},d]=\eta_p\frac{1}{\sqrt{k}}[\frac{1}{k-k_p}+\frac{1}{i(k+k_p)}] \label{pole}
\end{eqnarray}
with
$\eta_p=\frac{i\mu_\bot}{8\pi}\frac{k_p}{k_{1p}}\frac{e^{ik_{1p}h}e^{ik_{3p}d}}{\pi\sqrt{k_p}}\frac{N_{13}(k_p)}{\frac{\partial
D_{13}(k_p)}{\partial k_p}}$ that we will define as the SP field
$\tilde{\Psi}_{\textrm{SP}}[\mathbf{k},d]:=\tilde{\Psi}_{\textrm{TM}}^{pole}[\mathbf{k},d]$
(see~\cite{supll} for the general case). To justify this definition,
we point out that from Eq.~(\ref{pole}) we deduce a SP field
$\Psi_{\textrm{SP}}(\mathbf{x},z)=\int
d^{2}\mathbf{k}\tilde{\Psi}_{\textrm{TM}}^{pole}[\mathbf{k},d]e^{i\mathbf{k}\cdot\mathbf{x}}e^{ik_3(z-d)}$
which, in analogy with Eq.~(\ref{field}), is alternatively defined
by a contour integral over $\xi$ along $\Gamma$
\begin{eqnarray} \Psi_{\textrm{SP}}(\mathbf{x},z)=\int_{\Gamma}d\xi
F_{\textrm{SP}}(\xi)e^{ik_0nr\cos{(\xi-\vartheta)}}\label{fieldSP}
\end{eqnarray}
where
$F_{\textrm{SP}}(\xi)=\tilde{\Psi}_{\textrm{TM}}^{pole}[\mathbf{k},d]kk_3H_0^{(+)}(k|\mathbf{x}|)e^{-ik|\mathbf{x}|}$.
The remarkable fact about Eq.~(\ref{fieldSP}) is that it can be
evaluated by the procedures used for Eq.~(\ref{field}) without any
branch cut, and split into one contribution from the residue and
another from the SDP:
\begin{eqnarray}
\Psi_{\textrm{SP}}(\mathbf{x},z)&=&2\pi i
\textrm{Res}[F_{\textrm{SP}}(\xi_p)]e^{ik_0nr\cos{(\xi_p-\vartheta)}}\Theta(\vartheta-\vartheta_{\textrm{LR}})\nonumber\\
&+&e^{ik_0nr}\sum_{m\in
\textrm{even}}\frac{\Gamma(\frac{m+1}{2})}{m!(k_0nr)^{\frac{m+1}{2}}}\case{d^m}{d\tau^m}G_{\textrm{SP}}(0)\label{SPdeve}
\end{eqnarray}
with $G_{\textrm{SP}}(\tau)=F_{\textrm{SP}}(\xi)\frac{d\xi}{d\tau}$.
Importantly, the residue term in Eq.~(\ref{SPdeve}) is identical to
$\Psi_p$ given that $\textrm{Res}[F_{\textrm{SP}}(\xi_p)]=\textrm{Res}[F(\xi_p)]$.\\
\indent In the far field, the  $m=0$ term in the sum dominates and
we have  $\Psi_{\textrm{SP}}(\mathbf{x},z)\simeq\case{2\pi
k_0n\cos{\vartheta}}{ir}e^{ik_0
nr}\tilde{\Psi}_{\textrm{SP}}[\mathbf{k},d]$.  In Fig.~\ref{fig3} we
compare this expression for $\Psi_{\textrm{SP}}$ to
Eq.~(\ref{SDPm0}) by computing the intensity in the back-focal
plane. In the case of a vertical dipole -Fig.~\ref{fig3}(a)- the SP
term $I_{\textrm{SP},\Pi}[\mathbf{k}]$, proportional to
$|\tilde{\Psi}_{\textrm{SP}}[\mathbf{k},d]|^2$, is quasi-identical
to $I_{\textrm{TM},\Pi}[\mathbf{k}]$. We define the non-plasmonic
signal by $I_{\overline{\textrm{SP}}}\propto
|\tilde{\Psi}_{\textrm{TM}}[\mathbf{k},d]-\tilde{\Psi}_{\textrm{SP}}[\mathbf{k},d]|^2$.
In the case of a horizontal dipole -Fig.~\ref{fig3}(b)- there is
also an additional TE contribution
$I_{\textrm{TE},\Pi}[\mathbf{k}]\propto|\tilde{\Psi}_{\textrm{TE}}[\mathbf{k},d]|^2$
to $I_{\overline{\textrm{SP}}}$. The intensity dip observed for such
a horizontal dipole is attributed to a Fano-type interference effect
in the $\mathbf{k}$-space between the peaked SP contribution and the
broad non-plasmonic signal made explicit by our analysis~\cite{Sarrazin,Genet} .\\
\begin{figure}[htb] \centering
\includegraphics[width=8.6cm]{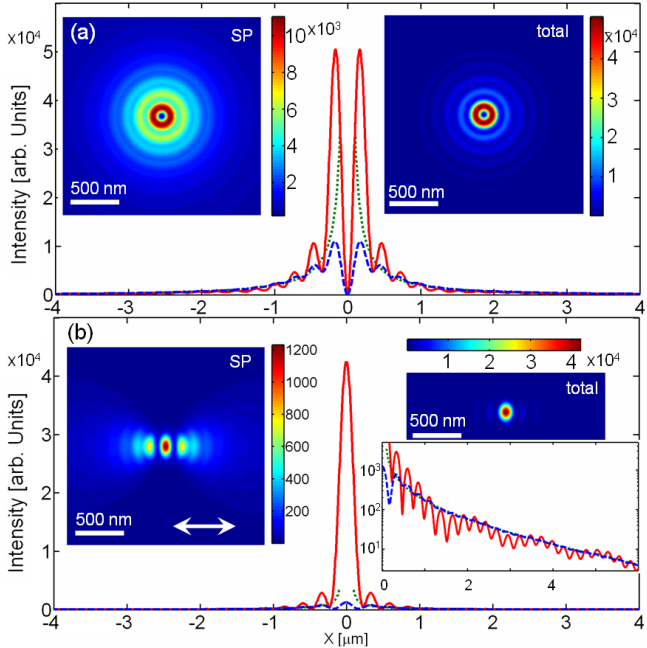}
\caption{Direct space images associated with a point-like dipole
radiating through the metal film and recorded in the back focal
plane of the microscope ocular. (a) Vertical dipole, (b) horizontal
dipole (including in this case the additional TE
contribution). The red curves correspond to the total signal computed from
Eqs.~(\ref{SDPm0}) and (\ref{paraSP}) and the blue ones to the mere
SP signal, i.e. Eqs.~(\ref{SPdeve}) and (\ref{paraSP}).}\label{fig4}
\end{figure}
\indent In a last step, we calculate direct space images through a
microscope ocular (see Fig.~\ref{fig1}(a)) associated with SP
propagation on the metal film by an inverse Fourier transform of the
field signal in the $\Pi$ plane, taking into account the finite
angular aperture of the objective~\cite{supll}. We compare in
Fig.~\ref{fig4} (a,b) the images calculated from Eqs.~(\ref{SDPm0})
and (\ref{SPdeve}) respectively for a vertical and a horizontal
dipole. Signal differences are more important for a horizontal
dipole where TE and TM fields interfere, and decrease for distances
larger than $2\pi/k_0$. We point out that the pure SP field at a
point $\mathbf{x}'$ of the image plane is given by a simple
expression
\begin{eqnarray}
\mathbf{E}_{\textrm{SP}}(\mathbf{x}')\propto \int_{|\mathbf{k}|\leq
k_0\textrm{NA}}d^{2}\mathbf{k}\sqrt{k_3}\mathbf{k}\tilde{\Psi}_{\textrm{SP}}[\mathbf{k},d]e^{i\frac{\mathbf{k}\cdot\mathbf{x'}}{M}}\nonumber\\
= -\int
d^2\mathbf{x}\mathbf{D}_{\textrm{SP},||}(\mathbf{x},d)\chi(\mathbf{x}+\frac{\mathbf{x'}}{M})/\sqrt{k_{3p}}  \label{paraSP}
\end{eqnarray}
where $\mathbf{D}_{\textrm{SP},||}(\mathbf{x},d)=\frac{\partial^2}{\partial\mathbf{x}\partial
z}\Psi_{\textrm{SP}}(\mathbf{x},d)$ is the in-plane component of the
SP displacement field along the interface  $z=d$ and
$\chi(\mathbf{u})\approx\case{k_0\textrm{NA}}{2\pi|\mathbf{u}|}J_1(k_0\textrm{NA}|\mathbf{u}|)
$ is the (scalar) point-spread function of the microscope objective. Taking a large microscope magnification
$M=nf'/f \ll 1$ (with $f,f'$ being respectively the objective and ocular focal lengths) enables us to analyze the recorded images simply
using the paraxial-like Eq.~(\ref{paraSP}), despite that leaky waves are emitted in a non-paraxial regime at
$\vartheta_{\textrm{LR}}$. Additionally, since
$\tilde{\Psi}_{\textrm{SP}}[\mathbf{k},d]|$ defines a sharp ring-like
distribution, we can approximately write
$\mathbf{E}_{\textrm{SP}}(\mathbf{x}')\propto\mathbf{D}_{\textrm{SP},||}(-\mathbf{x'}/M,d)$
for large $|\mathbf{x}'|/M$. Therefore, as shown in Fig.~\ref{fig4}, the
difference between the real image and the SP field vanishes
asymptotically when $|\mathbf{x}'|$ increases. Finally, this
analysis shows that only in-plane components of the SP field
participate to the image, therefore resolving definitively the current
controversy~\cite{WangOL2010,DePeraltaOL2010,HohenauOptX2011}.  \\
\indent To conclude, we have removed the long-standing ambiguity
associated with the very definition of a SP mode as probed in LRM
through a revision of the SP field. We have shown how this field
interferes, in a Fano-type way, with a broad non-plasmonic radiative
signal in the LR imaging process.  We expect our findings to have
important impact in the ever-growing field of plasmonics in its
different variants:
classical to quantum through molecular to non-linear plasmonics.\\
\indent This work was supported by Agence Nationale de la Recherche
(ANR), France, through the PLASTIPS project and by the french program Investiment d'Avenir (Equipex Union).


\begin{thebibliography}{99}
\bibitem{footLeg1} We model the high NA aplanatic immersion LRM objective by a reference sphere $\Sigma$ centered on $[x=0,y=0,z=0]$ and of radius  equal to the objective focal length $f$. The spherical wave front is transformed into a planar wave front $\Pi$ at the back focal plane of the objective and converges, after an ocular of focal $f^{\prime}$, in the image plane $\Pi^{\prime}$. Note that since
$\tan\vartheta_{\textrm{LR}}\geq 1$, non paraxial abberations
occurring during the propagation from $\Sigma$ to $\Pi$ have to be
accounted for~\cite{Sheppard}.
\bibitem{note1} The additional branch-cut implied in principle by the $1/\sqrt{k|\mathbf{x}|}$ term of the
Hankel function
can be chosen running just below the $\Gamma$ path with
therefore no influence on further contour evaluations.
\end{thebibliography}
\end{document}


\bibliographystyle{prsty}
\title{Supplementary for ``Imaging surface plasmons: the fingerprint of leaky waves on the far field''}
\author{Aur\'{e}lien Drezet, Cyriaque Genet}
\affiliation{Institut N\'eel UPR 2940, CNRS-University Joseph Fourier, 25 rue des Martyrs, 38000 Grenoble, France}
\affiliation{Laboratoire des Nanostructures, ISIS, Universit\'{e} de Strasbourg, CNRS (UMR7006) 8 all\'{e}e Gaspard Monge, 67000 Strasbourg, France}
\date{\today}

\pacs{42.25.Lc, 42.70.-a, 73.20.Mf} \maketitle
\section{introduction}
We start with a scalar potentials expansion of the electromagnetic
field in the three media $j=1,2,3$ corresponding respectively to air
metal and substrate (i.e. glass or fused silica). There is no source
in the substrate and the dipole (point source) is located in medium
$j=1$. We write for the field in each medium:
\begin{eqnarray}
\mathbf{D}_j=\boldsymbol{\nabla}\times\boldsymbol{\nabla}\times[\mathbf{\hat{z}}\Psi_{\textrm{TM},j}]+ik_0\varepsilon_j\boldsymbol{\nabla}\times[\mathbf{\hat{z}}\Psi_{\textrm{TE},j}]\nonumber\\
\mathbf{B}_j=\boldsymbol{\nabla}\times\boldsymbol{\nabla}\times[\mathbf{\hat{z}}\Psi_{\textrm{TE},j}]-ik_0\boldsymbol{\nabla}\times[\mathbf{\hat{z}}\Psi_{\textrm{TM},j}]\label{modale2}
\end{eqnarray} with
\begin{eqnarray}
[\boldsymbol{\nabla}^2+k_0^2\varepsilon_j]\Psi_{\textrm{TM,TE},j}=0.
\end{eqnarray}
Using Boundary conditions we show that the only non vanishing scalar
potentials for the dipole direction perpendicular to the film  is in
medium $j=3$
\begin{eqnarray}
\Psi^{\textrm{TM},\bot}(\mathbf{x},z)=\frac{i\mu_\bot}{4\pi}\int_{0}^{+\infty} \frac{kdk}{k_1}\tilde{T}_{13}^{\textrm{TM}}(k)e^{ik_1h}e^{ik_3z}J_0(k\varrho)\nonumber\\
=\frac{i\mu_\bot}{8\pi}\int_{-\infty}^{+\infty} \frac{kdk}{k_1}\tilde{T}_{13}^{\textrm{TM}}(k)e^{ik_1h}e^{ik_3z}H_0^{(+)}(k\varrho),
\end{eqnarray}
where we defined $k_i=\sqrt{k_0^2\varepsilon_i-k^2}$ with
$\textrm{Imag}[k_j]\geq0$, $k_0=2\pi/\lambda=\omega/c$, and $z\geq
d$. To obtain Eq.~3 we also used the formula
$H^{(+)}_0(u)-H^{(+)}_0(-u)=2J_0(u)$ which is valid in the complex
plane $u=u'+iu''$ (if $|\arg{(z)}|< \pi$). The Fresnel coefficient
characterizing the transmission of the metal film is for TM waves
defined by \begin{eqnarray}
\tilde{T}_{13}^{\textrm{TM}}(k)=\frac{T_{23}^{\textrm{TM}}T_{12}^{\textrm{TM}}}{1+R_{23}^{\textrm{TM}}R_{12}^{\textrm{TM}}e^{2ik_2d}}e^{i(k_2-k_3)d}
\end{eqnarray} where
\begin{eqnarray}
R_{ij}^{\textrm{TM}}=\frac{k_i/\varepsilon_i-k_j/\varepsilon_j}{k_i/\varepsilon_i+k_j/\varepsilon_j}\\
T_{ij}^{\textrm{TM}}=\frac{2k_i/\varepsilon_i}{k_i/\varepsilon_i+k_j/\varepsilon_j}.
\end{eqnarray}
We then introduce the variables $k=k_0n\sin{\xi}$, $k_3=k_0n\cos{\xi}$ with $\xi=\xi'+i\xi''$ and write
\begin{eqnarray}
\Psi^{\textrm{TM},\bot}(\mathbf{x},z)=\int_{\Gamma}d\xi F_+^{\textrm{TM},\bot}(\xi)e^{ik_0n((z-d)\cos{\xi}+\varrho\sin{\xi})}
\end{eqnarray}
with
\begin{eqnarray}
F_+^{\textrm{TM},\bot}(\xi)=\frac{i\mu_\bot}{8\pi}\frac{k_0n\sin{\xi}\cos{\xi}}{\sqrt{(\frac{\varepsilon_1}{\varepsilon_3}-\sin^2{\xi})}}
\tilde{T}_{13}^{\textrm{TM}}(k_0n\sin{\xi})\nonumber\\e^{i\sqrt{(\frac{\varepsilon_1}{\varepsilon_3}-\sin^2{\xi})}h}
\cdot e^{ik_0nd\cos{\xi}}H_0^{(+)}(k_0n\varrho\sin{\xi})e^{-ik_0n\varrho\sin{\xi}}\nonumber\\
\end{eqnarray} (we point out that the $\varrho$ and $\varphi$ dependencies are here and in the following implicit in our notation: $F_+(\xi):=F_+(\xi,\varphi,\varrho)$).
Similar expressions can be obtained for the components $\boldsymbol{\mu}_{||}$ of the dipole parallel  to the interface. More precisely  for the TM modes we have
\begin{eqnarray}
\Psi^{\textrm{TM},||}(\mathbf{x},z)=\int_{\Gamma}d\xi F_+^{\textrm{TM},||}(\xi)e^{ik_0n((z-d)\cos{\xi}+\varrho\sin{\xi})}
\end{eqnarray}
with
\begin{eqnarray}
F_+^{\textrm{TM},||}(\xi)=\frac{\boldsymbol{\mu}_{||}\cdot\hat{\boldsymbol{\varrho}}}{8\pi}k_0n\cos{\xi}
\tilde{T}_{13}^{\textrm{TM}}(k_0n\sin{\xi})\nonumber\\e^{i\sqrt{(\frac{\varepsilon_1}{\varepsilon_3}-\sin^2{\xi})}h}
\cdot e^{ik_0nd\cos{\xi}}H_1^{(+)}(k_0n\varrho\sin{\xi})e^{-ik_0n\varrho\sin{\xi}}.\nonumber\\
\end{eqnarray} Similarly for the TE waves we obtain:
\begin{eqnarray}
\Psi^{\textrm{TE},||}(\mathbf{x},z)=\int_{\Gamma}d\xi F_+^{\textrm{TE},||}(\xi)e^{ik_0n((z-d)\cos{\xi}+\varrho\sin{\xi})}
\end{eqnarray}
with\begin{eqnarray}
F_+^{\textrm{TE},||}(\xi)=-\frac{\boldsymbol{\mu}_{||}\cdot\hat{\boldsymbol{\varphi}}}{8\pi}\frac{k_0\cos{\xi}}{\sqrt{(\frac{\varepsilon_1}{\varepsilon_3}-\sin^2{\xi})}}
\tilde{T}_{13}^{\textrm{TE}}(k_0n\sin{\xi})\nonumber\\e^{i\sqrt{(\frac{\varepsilon_1}{\varepsilon_3}-\sin^2{\xi})}h}
\cdot e^{ik_0nd\cos{\xi}}H_1^{(+)}(k_0n\varrho\sin{\xi})e^{-ik_0n\varrho\sin{\xi}}.\nonumber\\
\end{eqnarray}We used the formula $H^{(+)}_1(u)+H^{(+)}_1(-u)=2J_1(u)$. Here the Fresnel coefficients are defined by \begin{eqnarray}
\tilde{T}_{13}^{\textrm{TE}}(k)=\frac{T_{23}^{\textrm{TE}}T_{12}^{\textrm{TE}}}{1+R_{23}^{\textrm{TE}}R_{12}^{\textrm{TE}}e^{2ik_2d}}e^{i(k_2-k_3)d}
\end{eqnarray} with
\begin{eqnarray}
R_{ij}^{\textrm{TE}}=\frac{k_i-k_j}{k_i+k_j}\\
T_{ij}^{\textrm{TE}}=\frac{2k_i}{k_i+k_j}.
\end{eqnarray}
The presence of the singular Hankel functions $H_1^{(+)}$ and
$H_0^{(+)}$ in all these expressions imply the existence of a branch
cut starting at the origin and associated with the function
$1/\sqrt{(\sin{\xi})}$. This branch cut is chosen in order to have
no influence during subsequent contour deformations and is running
just below the actual path $\Gamma$ slightly off the real axis
$\xi'$ and to the left of the vertical line $\xi''=-\pi/2$ (the
original branch cut is composed of the line $\xi'=-\pi/2$ and of the
half-axis [$\xi''=0$, $\xi'\leq 0$]). We also introduce the polar
coordinates $\varrho=r\sin{\vartheta}$, $z=d+r\cos{\vartheta}$
leading to $(z-d)\cos{\xi}+\varrho\sin{\xi}=r\cos{(\xi-\vartheta)}$
and therefore:
\begin{eqnarray}
\Psi(\mathbf{x},z)=\int_{\Gamma}d\xi F_+(\xi)e^{ik_0nr\cos{(\xi-\vartheta)}}.\label{field}
\end{eqnarray}
The definition of the square root
$k_1=k_0n\sqrt{(\frac{\varepsilon_1}{\varepsilon_3}-\sin^2{\xi})}$,
with $\varepsilon_3=n^2$ real and
$\varepsilon_1=\varepsilon_1'+i\varepsilon_1''\sim 1+i\delta$ with
$\delta\rightarrow 0^+$, implies the presence of a branch cut which
must be chosen carefully  in order i) to  be consistent with the
choice for $k_1$ made in Eq.~3 during integration along the contour
$\Gamma$, ii)  to allow further contour deformations leading to
convergent calculations. The branch cut adapted to our problem is
shown in Figs.~1,2 and correspond to the choice
$\textrm{Imag}[k_1]\geq0$ in the whole complex $\xi$-plane.  The
branch cut starts at the branch point  $M$ of coordinate $\xi_c$
defined by the condition $k_1=0$. We point out that due to
invariance of the Fresnel coefficient
$\tilde{T}_{13}^{\textrm{TM},\textrm{TE}}(k_0n\sin{\xi})$  over the
permutation $k_2\leftrightarrow- k_2$ we don't actually need an
additional branch cut for $k_2$ (this important property will
survive for a larger number of layers).
\section{The different contributions along the closed contour}
After introducing the function $f(\xi)=i\cos(\xi-\vartheta)=i\cos{(\xi'-\vartheta)}\cosh{\xi''}+\sin{(\xi'-\vartheta)}\sinh{\xi''}$ we define the steepest descent path $SDP$ by the condition\begin{eqnarray}\textrm{Imag}[f(\xi)]=\cos{(\xi'-\vartheta)}\cosh{\xi''}=1.~\label{SDP}\end{eqnarray} $SDP$ goes through the saddle point $\xi_0$ defined  by the condition $\frac{ df(\xi)}{d\xi}=0$ which has a solution at $\xi_0=\vartheta$. Importantly, there are actually two trajectories solutions of Eq.~\ref{SDP} and going through $\xi_0$. We choose the one  such that the real part of $f(\xi)$ decay uniformly along $SDP$ when going away arbitrarily to the left or to the right from the saddle point(see Fig.~1).\\
\indent Cauchy theorem allows us to deform the original $\Gamma$
contour to include $SDP$ as a part of the integration path. For this
we label $\Gamma$ by the letter $ABCD$ (see Fig.~1). The integral in
Eq.~\ref{field} is thus written $\int_{\Gamma}=\int_{ABCD}$. We will
consider two cases depending whether $\vartheta$ is or not larger
than the real part of the branch point $\xi'_c\simeq
\arcsin{(1/n)}=\vartheta_c$.
\subsection{Closing the contour in the case $\vartheta>\xi'_c$}
If $\vartheta>\xi'_c$ the closed integration contour contain eight
contributions (see Fig.~1) and we have:
\begin{equation}0=\int_{\Gamma}+\int_{DE}+\int_{EF}+\tilde{\int_{FG}}+\tilde{\int_{GH}}+\int_{HI}+\int_{IA}-I_{SP}.\end{equation}
The contribution \begin{eqnarray}
\int_{DE}:=\int_{DE}d\xi F_+(\xi)e^{ik_0nr\cos{(\xi-\vartheta)}}\nonumber \\
=\int_{\pi/2-i\infty}^{\pi/2+\vartheta-i\infty}d\xi F_+(\xi)e^{ik_0nr\cos{(\xi-\vartheta)}}
\end{eqnarray}
approaches zero asymptotically  and can therefore be neglected.\\
Similarly, we can neglect $\int_{IA}:=\int_{+i\infty}^{-\pi/2+i\infty}d\xi F_+(\xi)e^{ik_0nr\cos{(\xi-\vartheta)}}$ which approaches also zero asymptotically.\\
The contribution $\int_{EF}$ and $\tilde{\int_{FG}}$ are calculated along the $SDP$. However, due to the presence of the branch cut crossing $SDP$ at $F$ the integration along $FG$ actually corresponds to a change of Riemann sheet associated with the second determination for the square root $k_1$ (we point out that since the branch cut is very close to the imaginary axis at $F$ we have at the limit $\xi_F\simeq i\textrm{arccosh}{(1/\cos{\vartheta})}$). More precisely if  we call ``+'' the Riemann sheet in which $\textrm{Imag}[k_1]\geq0$ the second Riemann
\begin{figure*}[h!t]
\centering
\includegraphics[width=16cm]{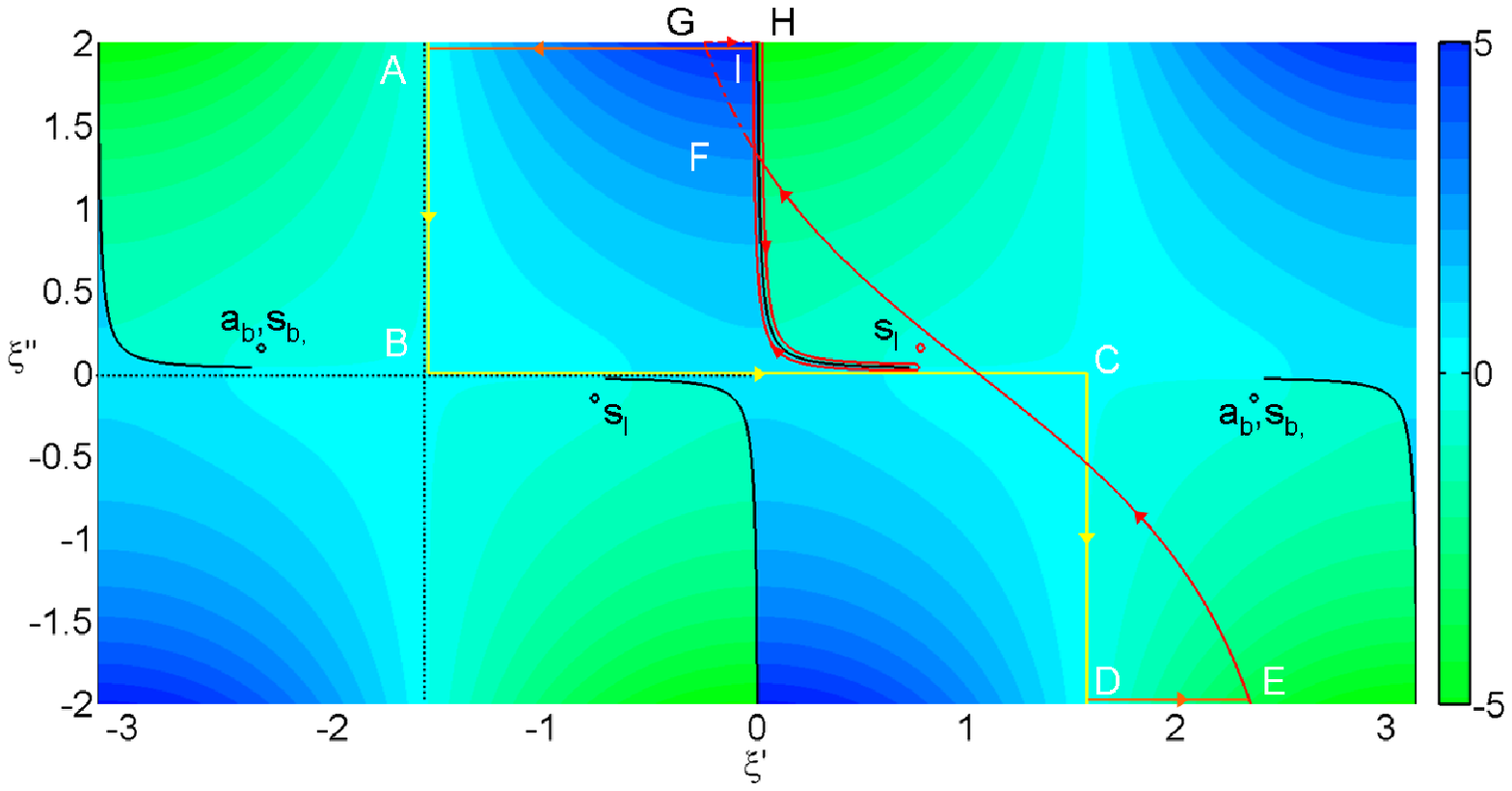}
\caption{Integration contour in the  complex $\xi$-plane for $\vartheta>\xi'_c$. }
\end{figure*}
surface ``-'' associated with the condition
$\textrm{Imag}[k_1]\leq0$ is connected to ``+'' through the branch
cut represented in Fig.~1. Therefore, crossing the branch cut at $F$
corresponds actually to a change in sign of the square root
$k_1\rightarrow-k_1$. We have consequently the contributions
\begin{eqnarray}
\int_{EF}:=\int_{EF}d\xi F_+(\xi)e^{ik_0nr\cos{(\xi-\vartheta)}}\nonumber \\
\tilde{\int_{FG}}:=\int_{FG}d\xi F_-(\xi)e^{ik_0nr\cos{(\xi-\vartheta)}}
\end{eqnarray}
where $F_{-}(\xi)$ is the same function of $k$ as $F_+(\xi)$ but $\sqrt{(\frac{\varepsilon_1}{\varepsilon_3}-\sin^2{\xi})}$ (defined with $\textrm{Imag}[\sqrt{(\frac{\varepsilon_1}{\varepsilon_3}-\sin^2{\xi})}]\geq 0$) is now replaced by  $-\sqrt{(\frac{\varepsilon_1}{\varepsilon_3}-\sin^2{\xi})}$.  More precisely the square root $z_{+}=\sqrt{g}$ of the complex variable $g'+ig''$ is defined on the ``+'' Riemann sheet by $z_{+}=\textrm{sign}(g'')\sqrt{((g'+|g|)/2)}+i\sqrt{((-g'+|g|)/2)}$ where $\textrm{sign}(x)=1$ if $x>0$, $\textrm{sign}(x)=-1$ if $x<0$ and $\textrm{sign}(x)=0$ if $x=0$. On the``-'' sheet we have therefore $z_{-}=-z_{+}$. An important remark concerns here the integration convergence along the SDP when approaching the vertical asymptotes $\pm\frac{\pi}{2}+\vartheta\mp i\infty$. Indeed, due to the presence of the coefficient $e^{\pm ik_0nh\sqrt{(\frac{\varepsilon_1}{\varepsilon_3}-\sin^2{\xi})}}$ in the definition of $F_{\pm}(\xi)$ it is not obvious that the integrand will take a finite value at infinity. Actually a careful study of the limit behaviour of $F_{\pm}(\xi)$ including the exponentials terms as
well as the Hankel function contribution shows that there is no convergence problem for $F_{+}(\xi)$ at infinity (this also explains why $\int_{DE}$ and $\int_{IA}$ goes to zero asymptotically). However when going on  the ``-'' Riemann sheet the convergence is not always ensured. We found that however no problem occurs on this second sheet as soon as the condition $z+\varrho\tan{\vartheta}>h$ is verified. In particular, no problem appears if we impose $z>h$. Since here we are interested in the asymptotic behavior valid for $z\gg h$ this condition will be automatically satisfied.\\
\indent This point is particularly relevant when we consider the contribution $\tilde{\int}_{GH}:=\int_{-\pi/2+\vartheta+i\infty}^{i\infty}d\xi F_-(\xi)e^{ik_0nr\cos{(\xi-\vartheta)}}$ which approaches zero if  the previous condition $z>h$ is fulfilled.
\indent From $H$ we thus cross the branch cut and go back to the ``+'' sheet. We thus obtain a contour $\int_{HI}=\int_{HI}d\xi F_+(\xi)e^{ik_0nr\cos{(\xi-\vartheta)}}$ longing the branch cut in the original ``+''  space and contourning the branch point $k_1=0$ (corresponding nearly to $\xi_c\simeq \arcsin{(1/n)}=\vartheta_c$). We will see in  the subsection D that this contribution corresponds to a lateral wave associated with a Goos-H\"{a}nchen effect in transmission.\\
\indent Finally, due to the presence of isolated singularities in the complex plane (i.e. poles) for the TM waves we must subtract a residue contribution $I_{SP}$ which value will precisely depends on the position $\vartheta$ along the real axis (i.e. whether or not the poles are encircled by the closed contour in the complex $\xi$-plane). A complete analysis of these singularities show that we can in principle extract from the transmission coefficient $\tilde{T}_{13}^{\textrm{TM}}(k)$ four poles corresponding to the four SP modes guided along the metal slab. However, the branch cut choice made here  allows only the existence of three solutions called respectively symmetric leaky ($s_l$), symmetric bound ($s_b$) and asymmetric bound ($a_b$) modes. The two bound modes are always well outside the region of integration and are never encircled by the contour.  Only the leaky mode $s_l$ can eventually contribute as a residue depending whether or not  the angle $\vartheta$ is larger than the leakage radiation angle $\vartheta_{LR}$ defined by the condition $\cos{(\xi_p'-\vartheta_{LR})}\cosh{\xi_p''}=1$ (with $\xi_p$ the complex coordinate of the SP pole $s_l$). This implies:
\begin{equation}
\vartheta_{LR}=\xi_p'+\arccos{(1/\cosh{\xi_p''})}\simeq \xi_p',
\end{equation}
and therefore the residue contribution is written:
\begin{eqnarray}
I_{SP}=2\pi i \textrm{Res}[F_+(\xi_p)]e^{ik_0nr\cos{(\xi_p-\vartheta)}}\Theta(\vartheta-\vartheta_{LR}).\nonumber\\
\end{eqnarray} In the following we write $k_p=k_0n\sin{\xi_p}$, $k_{3,p}=k_0n\cos{\xi_p}$ and $k_{1,p}=k_0n\sqrt{(\frac{\varepsilon_1}{\varepsilon_3}-\sin^2{\xi_p})}$ the pole wavevectors associated with this $s_l$ mode.
The calculation of the different residues is straightforward and leads  for the vertical dipole case to:
\begin{widetext}\begin{eqnarray}
\textrm{Res}[F_+^{\textrm{TM},\bot}(\xi_p)]e^{ik_0nr\cos{(\xi_p-\vartheta)}}=\frac{i\mu_\bot}{8\pi}\frac{k_0n\sin{\xi_p}\cos{\xi_p}}{\sqrt{(\frac{\varepsilon_1}{\varepsilon_3}-\sin^2{\xi_p})}}
\textrm{Res}[\tilde{T}_{13}^{\textrm{TM}}(k_0n\sin{\xi_p})]e^{i\sqrt{(\frac{\varepsilon_1}{\varepsilon_3}-\sin^2{\xi_p})}h}
\cdot e^{ik_0nz\cos{\xi_p}}H_0^{(+)}(k_0n\varrho\sin{\xi_p})\nonumber\\
\end{eqnarray}\end{widetext}
We now write $\tilde{T}_{13}^{\textrm{TM}}(k)$ as a rational fraction $\frac{N_{13}(k)}{D_{13}(k)}$ (with polynomial functions $N_{13}(k)$, $D_{13}(k)$ of the variable $k$) and therefore for the single pole $\xi_p$ we get  $$\textrm{Res}[\tilde{T}_{13}^{\textrm{TM}}(k_0n\sin{\xi_p})]=\frac{N_{13}(k_p)}{\frac{\partial D_{13}(k_0n\sin{\xi_p})}{\partial \xi_p} }=\frac{1}{k_{3,p}}\frac{N_{13}(k_p)}{\frac{\partial D_{13}(k_p)}{\partial k_p} }.$$ We thus have finally
\begin{eqnarray}
\textrm{Res}[F_+^{\textrm{TM},\bot}(\xi_p)]e^{ik_0nr\cos{(\xi_p-\vartheta)}}\nonumber\\
=\frac{i\mu_\bot}{8\pi}\frac{k_p}{k_{1,p}}e^{ik_{1,p}h}e^{ik_{3,p}z}\frac{N_{13}(k_p)}{\frac{\partial D_{13}(k_p)}{\partial k_p} }H_0^{(+)}(k_p\varrho).
\end{eqnarray}
A similar expression is obtained for the horizontal dipole:
 \begin{eqnarray}
\textrm{Res}[F_+^{\textrm{TM},||}(\xi_p)]e^{ik_0nr\cos{(\xi_p-\vartheta)}}\nonumber\\
=\frac{\boldsymbol{\mu}_{||}\cdot\hat{\boldsymbol{\varrho}}}{8\pi}e^{ik_{1,p}h}e^{ik_{3,p}z}\frac{N_{13}(k_p)}{\frac{\partial D_{13}(k_p)}{\partial k_p} }H_1^{(+)}(k_p\varrho).
\end{eqnarray}
There is no residue for the TE modes.\\
\indent Going back to the SDP contribution we define the variable $\tau=e^{i\pi/4}\sqrt{2}\sin{((\xi-\vartheta)/2)}$ which leads to $f(\xi)=i-\tau^2$. Along $SDP$ $\tau$ is real such that $\tau^2=-\sin{(\xi'-\vartheta)}\sinh{\xi''}\geq 0$. We thus obtain $\tau=2\sin{((\xi'-\vartheta)/2)}\cosh{(\xi''/2)}$. The saddle point corresponds to $\tau=0$ ($\tau>0$ if $\xi''<0$ and $\tau<0$ if $\xi''>0$ along $SDP$). With this new variable the point $F$ has therefore the coordinate $\tau_F\simeq-2\sin{(\vartheta/2)}\cosh{(\frac{1}{2}\textrm{arccosh}(1/\cos(\vartheta)))}=-2\sin{(\vartheta/2)}\sqrt{1+\frac{1}{\cos{\vartheta}}}<0$. Defining the term $I_{SDP}=-\int_{EF}-\tilde{\int_{FG}}$ we therefore obtain
\begin{eqnarray}
I_{SDP}=e^{ik_0nr}\{\int_{-\infty}^{\tau_F^-}d\tau G_{-}(\tau)e^{-k_0nr\tau^2}\nonumber\\+\int_{\tau_F^+}^{+\infty}d\tau G_{+}(\tau)e^{-k_0nr\tau^2}\}\nonumber\\
=e^{ik_0nr}\int_{-\infty}^{+\infty}d\tau\{G_{+}(\tau)[1-\Theta(\tau_F-\tau)]\nonumber\\+G_{-}(\tau)[1-\Theta(\tau-\tau_F)]\}e^{-k_0nr\tau^2}
\end{eqnarray} where we defined $G_{\pm}(\tau)=F_{\pm}(\xi)\frac{d\xi}{d\tau}$ and used $\frac{d\xi}{d\tau}=\sqrt{2}e^{-i\pi/4}/\cos{((\xi-\vartheta)/2)}$. We introduced the Heaviside step function $\Theta(x)$ defined as: $\Theta(x)=1$ if $x\geq0$ and $\Theta(x)=0$ otherwise. Importantly $\lim_{\tau\rightarrow \tau_F^+} G_{+}(\tau)=\lim_{\tau\rightarrow \tau_F^-}G_{-}(\tau)$ and therefore the function $G(\tau)=G_{+}(\tau)(1-\Theta(\tau_F-\tau))+G_{-}(\tau)(1-\Theta(\tau-\tau_F))$ which is not defined at $\tau_F$ can be prolonged without difficulties at $F$.
\subsection{Closing the contour in the case $\vartheta<\xi'_c$}
If $\vartheta<\xi'_c$ the closed integration contour contain 6
contributions (see Fig.~2) and we have:
\begin{equation}0=\int_{\Gamma}+\int_{DE}+\int_{EF}+\tilde{\int_{FG}}+\int_{GH}+\int_{HA}.\end{equation}
All these contribution but $\int_{NM}$ are defined on the ``+'' Riemann sheet. $\int_{HA}$ and $\int_{DE}$ tends asymptotically to zero for reasons already discussed in the previous \begin{figure*}[h!t]
\centering
\includegraphics[width=16cm]{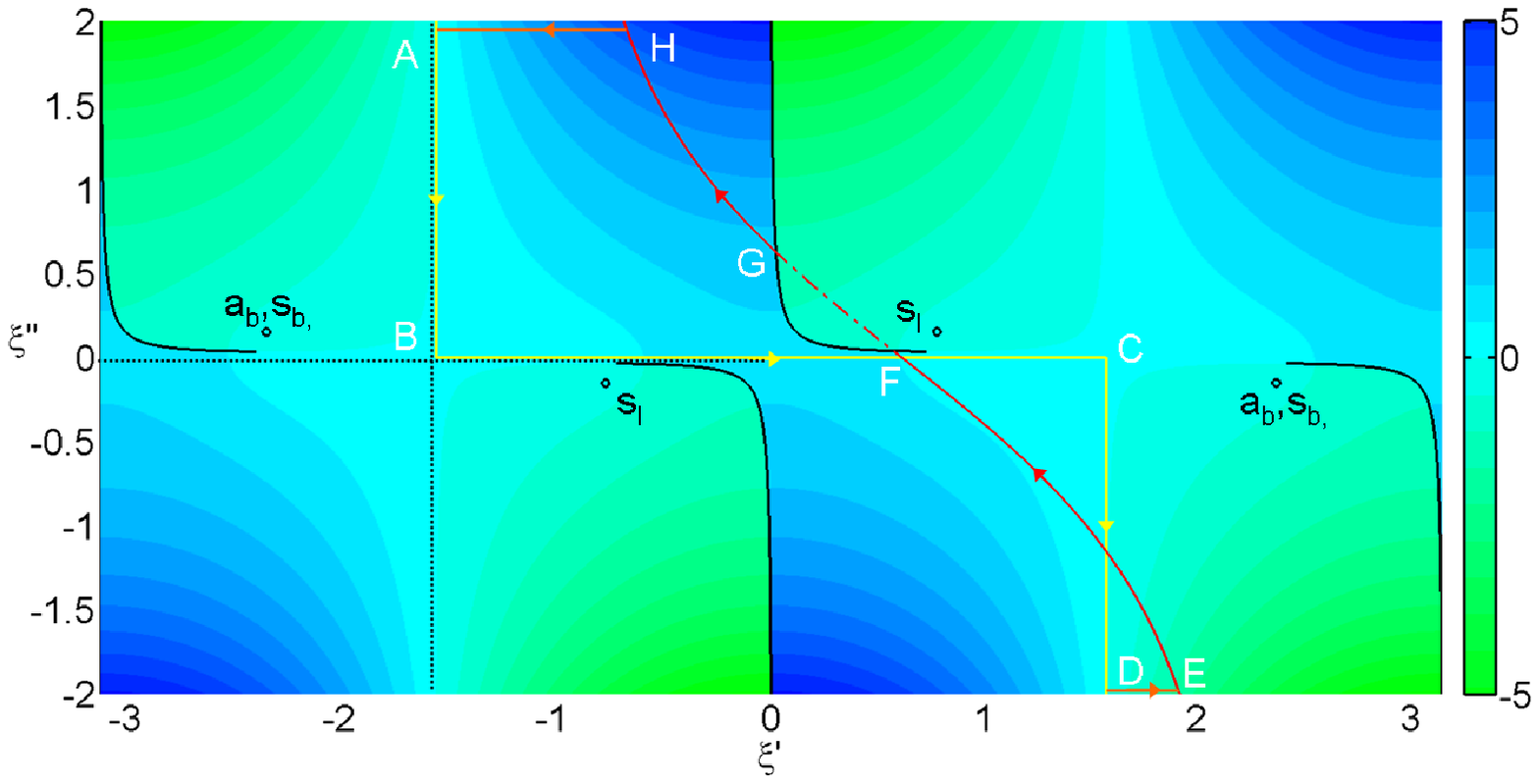}
\caption{Integration contour in the  complex $\xi$-plane for $\vartheta<\xi'_c$. }
\end{figure*}paragraph. Importantly there is no contribution along the branch cut since the integration path along SDP starts and finishes on the proper Riemann sheet ``+''. Regrouping the terms we thus have  for $\vartheta>\xi'_c$: $\int_{\Gamma}=I_{SDP}=-\int_{EF}-\tilde{\int_{FG}}-\int_{GH}$ with
\begin{eqnarray}
\int_{EF}:=\int_{EF}d\xi F_+(\xi)e^{ik_0nr\cos{(\xi-\vartheta)}}\nonumber \\
\tilde{\int_{FG}}:=\int_{FG}d\xi F_-(\xi)e^{ik_0nr\cos{(\xi-\vartheta)}}\nonumber \\
\int_{GH}:=\int_{GH}d\xi F_+(\xi)e^{ik_0nr\cos{(\xi-\vartheta)}}.
\end{eqnarray}
We then use the same variable $\tau$ and function $G_{\pm}(\tau)$ and thus obtain
\begin{eqnarray}
I_{SDP}=e^{ik_0nr}\{\int_{-\infty}^{\tau_F^-}d\tau G_{+}(\tau)e^{-k_0nr\tau^2}\nonumber\\+\int_{\tau_F^+}^{\tau_G^-}d\tau G_{+}(\tau)e^{-k_0nr\tau^2}\nonumber\\+\int_{\tau_G^+}^{-\infty}d\tau G_{+}(\tau)e^{-k_0nr\tau^2}\}\end{eqnarray}
which is rewritten as
\begin{eqnarray}
I_{SDP}=e^{ik_0nr}\int_{-\infty}^{\tau_F^-}d\tau G(\tau)e^{-k_0nr\tau^2}\end{eqnarray}
with\begin{eqnarray}
G(s)=G_{+}(\tau)[1-\Theta(\tau-\tau_F)]\nonumber\\
+G_{+}(\tau)[1-\Theta(\tau_G-\tau)]\nonumber\\+G_{-}(\tau)[1-\Theta(\tau_F-\tau)][1-\Theta(\tau-\tau_G)].
\end{eqnarray}
\subsection{The Steepest descent path contribution}
The previous integral $I_{SDP}$  for both $\vartheta>\xi'_c$ and $\vartheta<\xi'_c$ is of the gaussian form and can be evaluated  by doing a Taylor expansion of $G(\tau)$ around $\tau=0$. Using well known integrals we thus obtain
\begin{eqnarray}
I_{SDP}=e^{ik_0nr}\sum_{m\in \textrm{even}}\frac{\Gamma(\frac{m+1}{2})}{m!(k_0nr)^{\frac{m+1}{2}}}\frac{d^m}{d\tau^m}G(0).\label{taylor}
\end{eqnarray}
It is important to observe that $G(\tau)$ is highly singular in the vicinity of the SP pole $s_l$. Writing $\tau_p$ the coordinate of the pole in the $\tau$-space we thus define
\begin{eqnarray}
G(\tau):=G_0(\tau)+\frac{\textrm{Res}[G(\tau_p)]}{\tau-\tau_p}
\end{eqnarray} which (together with Eq.~\ref{taylor}) immediately implies
\begin{eqnarray}
I_{SDP}=e^{ik_0nr}\sum_{m\in \textrm{even}}\frac{\Gamma(\frac{m+1}{2})}{(k_0nr)^{\frac{m+1}{2}}}\{\frac{1}{m!}\frac{d^m}{d\tau^m}G_0(0)\nonumber\\
-\frac{\textrm{Res}[G(\tau_p)]}{\tau_p^{m+1}}\}.
\end{eqnarray}Remarkably, the singular integral $$I_{SDP}^{\textrm{pole}}:=e^{ik_0nr}\int_{-\infty}^{+\infty}d\tau\frac{\textrm{Res}[G(\tau_p)]}{\tau-\tau_p}e^{-k_0nr\tau^2}$$ can be directly calculated and we thus obtain
\begin{eqnarray}
I_{SDP}=e^{ik_0nr}\sum_{m\in \textrm{even}}\frac{\Gamma(\frac{m+1}{2})}{m!(k_0nr)^{\frac{m+1}{2}}}\frac{d^m}{d\tau^m}G_0(0)+I_{SDP}^{\textrm{pole}}\nonumber\\
\end{eqnarray}
with
\begin{eqnarray}
I_{SDP}^{\textrm{pole}}=-2i\pi\textrm{Res}[G(\tau_p)]e^{ik_0nr\cos{(\xi_p-\vartheta)}}\{\Theta(-\tau_p'')
\nonumber\\
-\frac{1}{2}\textrm{erfc}(-i\tau_p\sqrt{(k_0nr)})\}\nonumber\\
=-e^{ik_0nr}\sum_{m\in \textrm{even}}\frac{\Gamma(\frac{m+1}{2})}{(k_0nr)^{\frac{m+1}{2}}}\frac{\textrm{Res}[G(\tau_p)]}{\tau_p^{m+1}}
\end{eqnarray}
 where $\textrm{erfc}(z)=(2/\sqrt\pi)\int_{z}^{+\infty} e^{-t^2}dt$ is the Gauss complementary error function.
Notably, we have (see appendix) $\textrm{Res}[G(\tau_p)]=\textrm{Res}[F_{+}(\xi_p)]$ and  $\Theta(-\tau_p'')=\Theta(\vartheta-\vartheta_{LR})$ therefore $I_{SDP}^{\textrm{pole}}$ contains  up to the sign difference the same contribution which already appeared in  $I_{SP}$.  Consequently, the sum $I_{SDP}^{\textrm{pole}}+I_{SP}$ of the two contributions proportional to the residue represents a simple explicit mathematical expression:
\begin{eqnarray}
I_{SDP}^{\textrm{pole}}+I_{SP}\nonumber\\=i\pi\textrm{Res}[G(\tau_p)]e^{ik_0nr\cos{(\xi_p-\vartheta)}}\textrm{erfc}(-i\tau_p\sqrt{(k_0nr)}).
\end{eqnarray}
This sum is sometimes by definition associated with the surface plasmon mode. We point out however that the error function is highly singular and therefore we should preferably use the equivalent expression:
\begin{eqnarray}
I_{SDP}^{\textrm{pole}}+I_{SP}=2i\pi\textrm{Res}[G(\tau_p)]e^{ik_0nr\cos{(\xi_p-\vartheta)}}\Theta(\vartheta-\vartheta_{LR})\nonumber\\
-e^{ik_0nr}\sum_{m\in \textrm{even}}\frac{\Gamma(\frac{m+1}{2})}{(k_0nr)^{\frac{m+1}{2}}}\frac{\textrm{Res}[G(\tau_p)]}{\tau_p^{m+1}}.\nonumber\\
\end{eqnarray}We also note that most of the discussions and confusions made during the XX$^{th}$ on the role of SPs in the Sommerfeld integral  resulted from the above mentioned intricate relationship existing between the two singular terms $I_{SP}$ and  $I_{SDP}^{\textrm{pole}}$. For a historical discussion see Collin\cite{Collin}.
\subsection{The lateral wave contribution: Goos-H\"{a}nchen effect in transmission}
In the case $\vartheta>\xi'_c$ the integral $\int_{HI}$ along the
branch cut can be transformed using the method described in Ref.~2.
For this we separate the integral $\int_{HI}d\xi
F_+(\xi)e^{ik_0nr\cos{(\xi-\vartheta)}}$ into
\begin{figure*}[h!t]
\centering
\includegraphics[width=12cm]{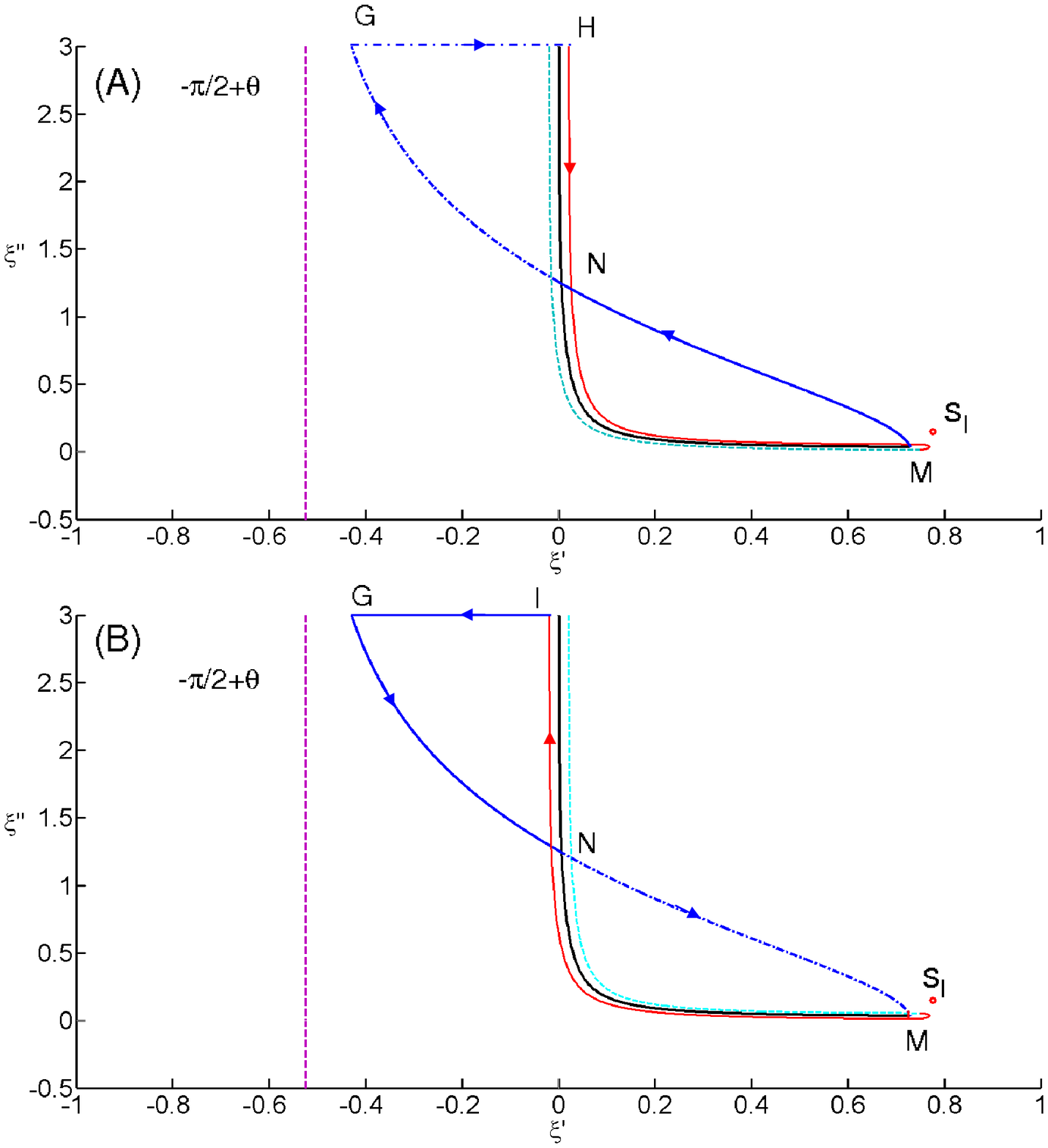}
\caption{Integration contour in the  complex $\xi$-plane  along the branch cut $HI$ around the branch point $M$ for $\vartheta>\xi'_c$. (A) shows the closed contour used to deform analytically the contour $HM$. (B) shows the  closed contour used to deform the contour $MI$.   }
\end{figure*}
a contribution $\int_{HM}=\int_{HM}d\xi
F_+(\xi)e^{ik_0nr\cos{(\xi-\vartheta)}}$ starting at  infinity at
$\xi=i\infty+0^+$ and stopping at the branch-point $M$ ($\xi_c\simeq
\vartheta_c$) and into a contribution $\int_{MI}=\int_{MI}d\xi
F_+(\xi)e^{ik_0nr\cos{(\xi-\vartheta)}}$ starting at $M$ and
finishing  at infinity $\xi=i\infty+0^-$ on the other side of the
branch cut. As shown in Fig.~3(A)  in order to calculate $I_{HM}$
the integration contour is closed by longing the modified steepest
descent path $MG$ defined by the equation
\begin{eqnarray}
\cos{(\xi'-\vartheta)}\cosh{\xi''}=\cos{(\xi_c'-\vartheta)}\cosh{\xi_c''}\nonumber\\ \simeq\cos{(\vartheta_c-\vartheta)}=K<1.
\end{eqnarray}
The curve $MG$  with a vertical asymptote at $\xi=-\pi/2+\vartheta$
is thus defined by
$\xi''=\textrm{arccosh}(K/\cos{(\xi'-\vartheta)})$. We thus have
$0=\int_{HM}+\int_{MN}+\tilde{\int}_{NG}+\tilde{\int}_{GH}$ where
$N$ is the intersection point between the modified steepest descent
path $MG$ and the branch cut ($\xi_N\simeq
i\textrm{arccosh}(K/\cos{(\vartheta)})$).
$\tilde{\int}_{NG}:=\int_{NG}d\xi
F_-(\xi)e^{ik_0nr\cos{(\xi-\vartheta)}}$ and
$\tilde{\int}_{GH}:=\int_{GH}d\xi
F_-(\xi)e^{ik_0nr\cos{(\xi-\vartheta)}}$ are evaluated on the ``-''
Riemann sheet while  $\int_{HM}:=\int_{HM}d\xi
F_+(\xi)e^{ik_0nr\cos{(\xi-\vartheta)}}$ and
$\int_{MN}:=\int_{MN}d\xi F_+(\xi)e^{ik_0nr\cos{(\xi-\vartheta)}}$
are evaluated on the ``+'' Riemann sheet. From $H$ we cross a second
time the branch cut in order to close the contour on the ``+''
Riemann sheet.\\
A similar analysis is done for the integration contour $\int_{MI}$.
We have $0=\int_{MI}+\int_{IG}+\int_{GN}+\tilde{\int}_{NM}$ where
$\int_{MI}$, $\int_{IG}$ and $\int_{GN}$ are defined as previously
on  the ``+'' Riemann sheet while $\tilde{\int}_{NM}:=\int_{NM}d\xi
F_-(\xi)e^{ik_0nr\cos{(\xi-\vartheta)}}$ is evaluated on the ``-''
Riemann sheet. In order to close the contour in ``+''  we must
finally cross the branch cut in the region surrounding $M$. The
infinitesimal  loop surrounding  $M$ gives however a vanishing
contribution which can be neglected.\\ Regrouping all these
expressions we define $I_{LW}=-\int_{HM}-\int_{MI}$ and we obtain
\begin{widetext}\begin{eqnarray}
I_{LW}=\int_{\xi_c}^{\xi_N} d\xi [F_+(\xi)-F_+(\xi)]e^{ik_0nr\cos{(\xi-\vartheta)}}
+\int_{\xi_N}^{-\pi/2+\vartheta+i\infty} d\xi [F_-(\xi)-F_+(\xi)]e^{ik_0nr\cos{(\xi-\vartheta)}}+\int_{IG}+\tilde{\int_{GH}}.
\end{eqnarray}\end{widetext}
The contributions $\int_{IG}$, $\tilde{\int_{GH}}$ vanish asymptotically as discussed before and therefore can be neglected.
Importantly due to the definition of the square root the function $F_-(\xi)-F_+(\xi)$ tends to vanish at the intersection point $N$.
We then define the function $\Phi(\xi)=[F_+(\xi)-F_+(\xi)]\textrm{sign}(\xi''_N-\xi'')$  vanishing at $\xi_c$ and write
\begin{eqnarray}
I_{LW}=\Theta(\vartheta-\xi'_c)\int_{\xi_c}^{-\pi/2+\vartheta+i\infty} d\xi \Phi(\xi)e^{ik_0nr\cos{(\xi-\vartheta)}}.\nonumber\\
\end{eqnarray} The Heaviside function was introduced in order to remember that $I_{LW}$ is only defined if $\xi'_c<\vartheta$.
In the present work we will only evaluate $I_{LW}$ approximately
using the method discussed in Ref.~2.  First, we observe that
$e^{ik_0nr\cos{(\xi-\vartheta)}}=e^{ik_0nrK}e^{k_0nr\sin{(\xi'-\vartheta)}\sinh{\xi''}}$.
Second, considering that only $\xi$ values in the vicinity of
$\xi_c\simeq\vartheta_c$ contribute significantly to  $I_{LW}$ we
write $d\xi\approx id\xi''$, and
$\sin{(\xi'-\vartheta)}\sinh{\xi''}\approx-\sin{(\vartheta-\vartheta_c)}\xi''<0$.
We therefore obtain
\begin{widetext}\begin{eqnarray}
I_{LW}\approx ie^{ik_0nrK}\Theta(\vartheta-\vartheta_c)\int_{0}^{+\infty} d\xi'' \Phi(\vartheta_c+i\xi'')e^{-k_0nr\sin{(\vartheta-\vartheta_c)}\xi''}\nonumber\\
=2ie^{ik_0nrK}\Theta(\vartheta-\vartheta_c)\int_{0}^{+\infty} udu \Phi(\vartheta_c+iu^2)e^{-k_0nr\sin{(\vartheta-\vartheta_c)}u^2}
\end{eqnarray}\end{widetext} where we used the variable $\xi''=u^2$. This integral is of the Gaussian kind and can be computed exactly using a Taylor expansion of $\Phi$ near $\vartheta_c$.  We consequently deduce
\begin{widetext}\begin{eqnarray}
I_{LW}\approx ie^{ik_0nrK}\Theta(\vartheta-\vartheta_c)\sum_{m=1}^{+\infty}\frac{\Gamma(1+m/2)}{(k_0nr\sin{(\vartheta-\vartheta_c)})^{1+m/2}}\frac{H^{(m)}(0)}{m!}
\end{eqnarray}\end{widetext} where we used the series expansion $\Phi(\vartheta_c+iu^2)=H(u)=\sum_{m=1}^{+\infty}\frac{u^m}{m!}\frac{d^m}{du^m}H(u)|_{u=0}=\sum_{m=1}^{+\infty}u^m\frac{H^{(m)}(0)}{m!}$ (the term $m=0$ vanishes since $\Phi(\theta_c)=0$).\\
\begin{figure}[h]
\centering
\includegraphics[width=8.2cm]{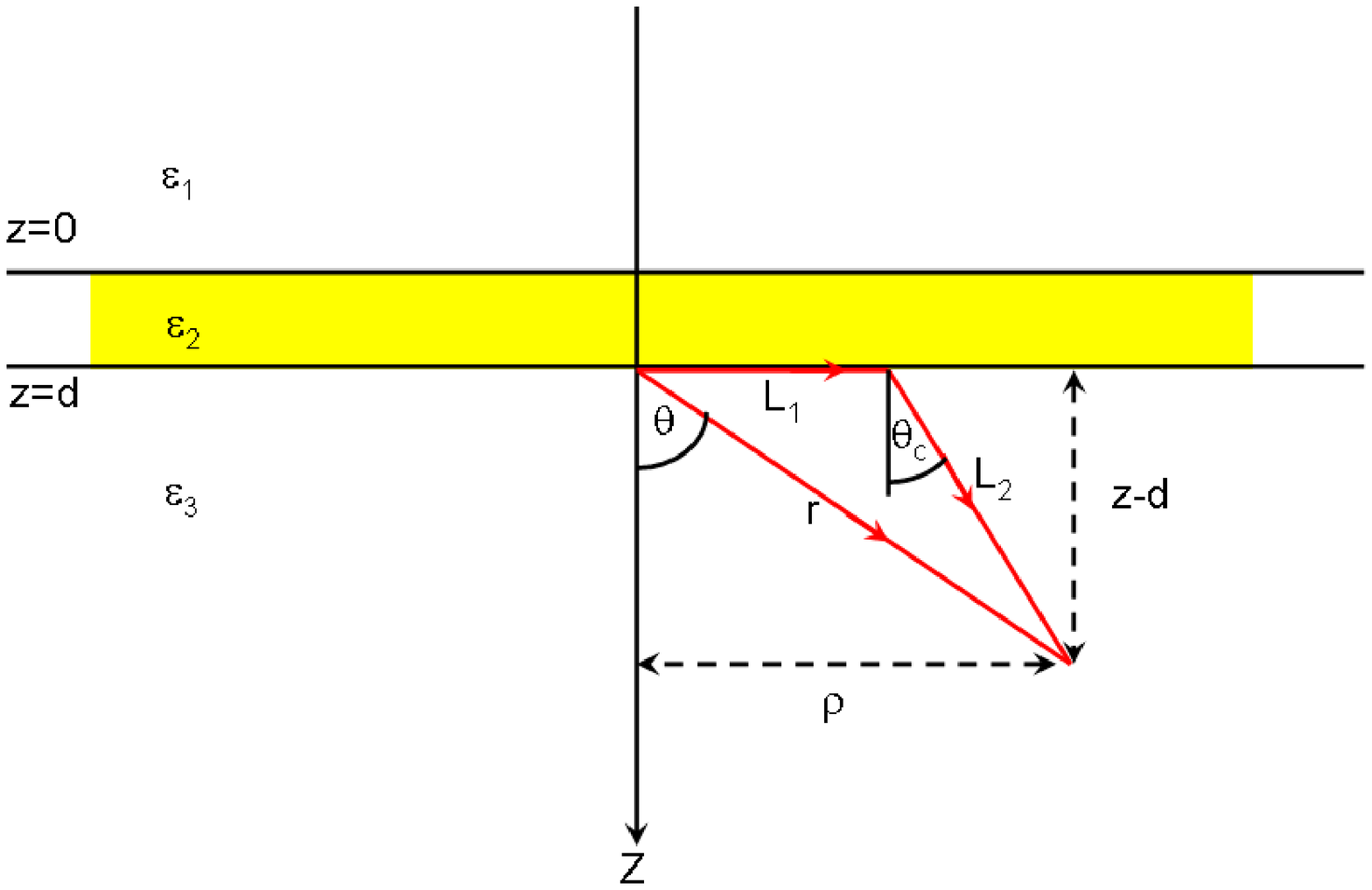}\caption{Geometric construction of the Goos-H\"{a}nchen phase in transmission.}
\end{figure}
The phase
$\delta \varphi = k_0nr\cos{(\vartheta-\vartheta_c)}=k_0nr[\cos{\vartheta}\cos{\vartheta_c}+\sin{\vartheta}\sin{\vartheta_c}]$ takes a simple interpretation if you define the length $L_1$ of $L_2$ by:
\begin{eqnarray}
r\sin{\vartheta}=\varrho=L_1+L_2\sin{\vartheta_c}\nonumber\\
r\cos{\vartheta}=z-d=L_2\cos{\vartheta_c}.
\end{eqnarray}
Therefore we obtain
\begin{eqnarray}
\delta \varphi = k_0n(L_2+L_1\sin{\vartheta_c})=k_0nL_2+k_0L_1
\end{eqnarray}
where we used $n\sin{\vartheta_c}=1$. As it is clear from Fig.~4 $L_1$ is the path length of a `creeping' wave propagating along the interface before to be re-emitted at the critical angle $\vartheta_c$. The re-emitted waves propagates along a distance $L_2$ in the medium of optical index $n$ and then reaches the point defined by the coordinates $(r,\vartheta)$.  The phase $\delta\varphi$ is thus generated by a virtual propagation of length $L_1$ along the interface air-dielectric $z=d$  (supposing no metal is present and that the volume corresponding to the film  between $z=0$ and $z=d$ is filled with the medium  of permittivity $\varepsilon_1\simeq 1$) and followed by a re-emission at the critical angle in the glass substrate $\varepsilon_3=n^2$. The previous analysis justifies therefore the name ``lateral'' we gave to the contribution $I_{LW}$. This effect can be seen as a kind of Goos-H\"{a}nchen deflection in transmission and is somehow equivalent to the already known Goos-H\"{a}nchen effect associated with lateral waves in the reflection mode.\\
\subsection{The Far-field Fraunhofer regime}
We are interested into evaluating the different integrals when $r\rightarrow +\infty$.
As a first approximation, concerning $I_{SDP}$ we calculate  only the term $m=0$ in the sum which reads:
\begin{eqnarray}
I_{SDP,m=0}=\frac{\sqrt{\pi}e^{ik_0 nr}}{\sqrt{k_0nr}}G(0)=\frac{\sqrt{2\pi}e^{ik_0 nr}e^{-i\pi/4}}{\sqrt{k_0nr}}F_+(\vartheta).
\end{eqnarray}
In the far-field where $r>>\lambda$ the Hankel  function can be approximated using the asymptotic formulas
\begin{eqnarray}
H^{(+)}_0(x)=\sqrt{\frac{2}{\pi x}}e^{-i\pi/4}(1-\frac{i}{8x})e^{ix}+O(x^{-5/2})\nonumber\\
H^{(+)}_1(x)=\sqrt{\frac{2}{\pi x}}e^{-i3\pi/4}(1+\frac{3i}{8x})e^{ix}+O(x^{-5/2})
\end{eqnarray} which are valid for $x>>1$.
Therefore for the vertical dipole we get
 \begin{widetext}\begin{eqnarray}
I_{SDP,m=0}^\bot=\frac{2\pi k_0n\cos{\vartheta}}{ir}e^{ik_0 nr}\tilde{\Psi}^{\textrm{TM},\bot}[k_0n\sin{\vartheta}\boldsymbol{\hat{\varrho}},z=d(1-\frac{i}{8k_0nr\sin{\vartheta}^2}+...)
\end{eqnarray}\end{widetext} where \begin{eqnarray}\tilde{\Psi}_{\textrm{TM},\bot}[\mathbf{k},z=d]=\frac{i\mu_\bot}{8\pi^2 k_1}\tilde{T}_{13}^{\textrm{TM}}(k)e^{ik_3 d}e^{ik_1 h}\nonumber\\=\frac{i\mu_\bot}{8\pi^2 k_0\sqrt{(1-n^2\sin{\vartheta}^2)}}\tilde{T}_{13}^{\textrm{TM}}(k)e^{ik_3 d}e^{ik_1 h}\end{eqnarray} is the 2D Fourier transform of  $\Psi^{\textrm{TM},\bot}(\varrho,z)$ calculated at $z=d$ (i.e $\int \frac{d^2\mathbf{x}}{4\pi^2}\Psi^{\textrm{TM},\bot}(\varrho,z=d)e^{-i\mathbf{k}\cdot\mathbf{x}}$) for the wavevector $\mathbf{k}=k_0n\sin{\vartheta}\boldsymbol{\hat{\varrho}}$.
Similarly for the Horizontal dipole we obtain for TM components:
\begin{widetext}\begin{eqnarray}
I_{SDP,m=0}^{||}=\frac{2\pi k_0n\cos{\vartheta}}{ir}e^{ik_0 nr}\tilde{\Psi}^{\textrm{TM},||}[k_0n\sin{\vartheta}\boldsymbol{\hat{\varrho}},z=d](1+\frac{3i}{8k_0nr\sin{\vartheta}^2}+...)
\end{eqnarray}\end{widetext}
where\begin{eqnarray}\tilde{\Psi}^{\textrm{TM},||}[\mathbf{k},z=d]=\frac{-i\mu_{||}\cdot\mathbf{k}}{8\pi^2 k^2}\tilde{T}_{13}^{\textrm{TM}}(k)e^{ik_3 d}e^{ik_1 h}\nonumber\\=\frac{-i\mu_{||}\cdot\boldsymbol{\hat{\varrho}}}{8\pi^2 k_0nr\sin{\vartheta}}\tilde{T}_{13}^{\textrm{TM}}(k)e^{ik_3 d}e^{ik_1 h}.\end{eqnarray}For  TE components we have also:
\begin{widetext}\begin{eqnarray}
I_{SDP,m=0}^{||}=\frac{2\pi k_0n\cos{\vartheta}}{ir}e^{ik_0 nr}\tilde{\Psi}^{\textrm{TE},||}[k_0n\sin{\vartheta}\boldsymbol{\hat{\varrho}},z=d](1+\frac{3i}{8k_0nr\sin{\vartheta}^2}+...)
\end{eqnarray}\end{widetext} with now
\begin{eqnarray}\tilde{\Psi}^{\textrm{TE},||}[\mathbf{k},z=d]=\frac{ik_0\mu_{||}\cdot(\mathbf{\hat{z}}\times\mathbf{k})}{8\pi^2 k_1k^2}\tilde{T}_{13}^{\textrm{TE}}(k)e^{ik_3 d}e^{ik_1 h}\nonumber\\=\frac{i\mu_{||}\boldsymbol{\hat{\varphi}}}{8\pi^2 k_0 \sqrt{(1-n^2\sin{\vartheta}^2)}n\sin{\vartheta}}\tilde{T}_{13}^{\textrm{TE}}(k)e^{ik_3 d}e^{ik_1 h}.\end{eqnarray}
In the far field only the term in $1/r$ survives and (in agreement
with the Stratton-Chu formalism~\cite{stratton} and Richards and
Wolf~\cite{wolf})~we can always write:
\begin{eqnarray}
\Psi\simeq I_{SDP,m=0}\simeq\frac{2\pi k_0n\cos{\vartheta}}{ir}e^{ik_0nr}\nonumber\\ \tilde{\Psi}^{\textrm{TM or te}}[k_0n\sin{\vartheta}\boldsymbol{\hat{\varrho}},z=d].\end{eqnarray}
\subsection{The intermediate regime: Generalization of the Norton wave }
The next term in the power expansion of $\Psi$ contributes proportionally to $1/r^2$.   To evaluate this term we must take into account not only $I_{SDP,m=0}$ but also $I_{SDP,m=2}$ and $I_{LW,m=1}$.
We use the notation \begin{eqnarray}F_{\pm}(\xi)=\sqrt{(\frac{2\pi k_0n}{r})}e^{-i\frac{\pi}{4}}k_0n\cos{\xi}Q_{\pm}^{\alpha}(k_0n\sin{\xi})\nonumber\\ \cdot[1-i\frac{1-4\alpha^2}{k_0nr(\sin{\xi})^2}+...]\end{eqnarray}  (with $\alpha=0$ or 1 depending whether the dipole is vertical or horizontal) and we obtain for the SDP contributions proportional to $1/r^2$:
\begin{eqnarray}I_{SDP,m=0}=\frac{2\pi(1-4\alpha^2)}{r^2}e^{ik_0nr}\frac{\cos{\vartheta}Q_{+}^{\alpha}(k_0n\sin{\xi})}{(\sin{\vartheta})^2}\end{eqnarray}
and
\begin{eqnarray}
I_{SDP,m=2}=\frac{e^{ik_0nr}\sqrt{\pi}}{4(k_0nr)^{3/2}}\frac{d^2G(\tau)}{d\tau^2}|_{\tau=0}\nonumber\\
=\frac{-\pi}{r^2}e^{ik_0nr}\frac{d^2[\frac{\cos{\xi}}{\cos{((\xi-\vartheta)/2)}}Q_{+}^{\alpha}(k_0n\sin{\xi})]}{d\xi^2}|_{\xi=\vartheta}.\nonumber\\
\end{eqnarray}
We also have to include the lateral wave (i.e. Goos-H\"{a}nchen) contribution:\begin{eqnarray}I_{LW,m=1}\simeq e^{ik_0nrK}\frac{i\sqrt{\pi}\Theta(\vartheta-\vartheta_c)}{2(k_0nr\sin{(\vartheta-\vartheta_c)})^{3/2}}\frac{d H(u)}{du}|_{u=0}\end{eqnarray} which reads
\begin{widetext}\begin{eqnarray}I_{LW,m=1}= \frac{\pi e^{ik_0(nL_2+L_1)}\Theta(\vartheta-\vartheta_c) e^{i\frac{\pi}{4}}}{r^2(\sin{(\vartheta-\vartheta_c)})^{3/2}}\frac{d\{\cos(\vartheta_c+iu^2)[Q_{+}^{\alpha}(k_0n\sin{(\vartheta_c+iu^2)})-Q_{-}^{\alpha}(k_0n\sin{(\vartheta_c+iu^2)})]\}}{du}|_{u=0}.
\end{eqnarray}\end{widetext}
The sum $I_{SDP,m=0}+I_{SDP,m=2}+I_{LW,m=1}$ describes an asymptotic field varying as $1/r^2$ and which constitutes a generalization of the result obtained by Norton for the radio wave antenna on a conducting earth problem.
\section{How to define the surface plasmon mode?}
\subsection{From the near-field to the far-field}
As seen in Section 2.E  the dominant contribution in the far-field has the form
\begin{eqnarray}
\Psi(\mathbf{x},z)=\frac{2\pi k_0n\cos{\vartheta}}{ir}e^{ik_0 nr}\tilde{\Psi}[k_0n\sin{\vartheta}\boldsymbol{\hat{\varrho}},z=d].
\end{eqnarray}
From Eq.~53 we also have the relation
\begin{eqnarray}\tilde{\Psi}[k_0n\sin{\vartheta}\boldsymbol{\hat{\varrho}},z=d]:=Q(s)\nonumber\\=\sqrt{(\frac{r}{2\pi k_0n})}e^{i\frac{\pi}{4}}\frac{F_+(\xi)}{k_0n\cos{\xi}}\nonumber\\=\sqrt{(\frac{r}{2\pi k_0n})}e^{i\frac{\pi}{4}}F_+(\xi)\frac{d\xi}{ds}\end{eqnarray} where $s=k_0n\sin{\xi}$ and where $\xi$ is here identical to $\vartheta$ (as usual the $\varrho$ and $\varphi$ dependencies are here implicit in $Q(s):=Q(s,\varphi,\varrho)$ and $F_+(\xi):=F_+(\xi,\varphi,\varrho)$).
In the  complex plane $\xi=\xi'+i\xi''$  and $s=s'+is''$ we have the singular/regular decomposition: $Q(s)=Q_0(s)+\textrm{Res}[Q(s_p)]/(s-s_p)$. Furthermore, from Eq.~59 this implies
\begin{eqnarray}
\frac{1}{2\pi i}\oint_{C_p}ds Q(s)=\textrm{Res}[Q(s_p)]\nonumber\\
=\sqrt{(\frac{r}{2\pi k_0n})}e^{i\frac{\pi}{4}}\frac{1}{2\pi i}\oint_{\mathcal{C}_p}d\xi F_+(\xi)
\nonumber\\
=\sqrt{(\frac{r}{2\pi k_0n})}e^{i\frac{\pi}{4}}\textrm{Res}[F_+(\xi_p)]
\end{eqnarray}
where $C_p$ and $\mathcal{C}_p$ are small closed contours surrounding the plasmon pole in respectively the complex $s$-plane and $\xi$-plane. Therefore, we can equivalently write
\begin{eqnarray}
Q(s)=Q_0(s)+\sqrt{(\frac{r}{2\pi k_0n})}e^{i\frac{\pi}{4}}\frac{\textrm{Res}[F_+(\xi_p)]}{s-s_p}.
\end{eqnarray}
The calculations being done in the far-field limit, where $r,\varrho\rightarrow +\infty$, we have for the vertical dipole case the residue:
\begin{widetext}\begin{eqnarray}
\textrm{Res}[F_+^{\textrm{TM},\bot}(\xi_p)]
=\frac{i\mu_\bot}{8\pi}\frac{k_p}{k_{1,p}}e^{ik_{1,p}h}e^{ik_{3,p}d}\frac{N_{13}(k_p)}{\frac{\partial D_{13}(k_p)}{\partial k_p} }H_0^{(+)}(k_p\varrho)e^{-ik_{p}\varrho}
\simeq\frac{i\mu_\bot}{8\pi}\frac{k_p}{k_{1,p}}e^{ik_{1,p}h}e^{ik_{3,p}d}\frac{N_{13}(k_p)}{\frac{\partial D_{13}(k_p)}{\partial k_p} }\sqrt{(\frac{2}{\pi k_p \varrho})}e^{-i\frac{\pi}{4}},2
\end{eqnarray}\end{widetext}
and similarly for the horizontal dipole residue:
\begin{widetext}\begin{eqnarray}
\textrm{Res}[F_+^{\textrm{TM},||}(\xi_p)]=\frac{\boldsymbol{\mu}_{||}\cdot\hat{\boldsymbol{\varrho}}}{8\pi}e^{ik_{1,p}h}e^{ik_{3,p}d}\frac{N_{13}(k_p)}{\frac{\partial D_{13}(k_p)}{\partial k_p} }H_1^{(+)}(k_p\varrho)e^{-ik_{p}\varrho}
\simeq\frac{\boldsymbol{\mu}_{||}\cdot\hat{\boldsymbol{\varrho}}}{8\pi}e^{ik_{1,p}h}e^{ik_{3,p}d}\frac{N_{13}(k_p)}{\frac{\partial D_{13}(k_p)}{\partial k_p} }\sqrt{(\frac{2}{\pi k_p \varrho})}e^{-i\frac{3\pi}{4}}.
\end{eqnarray}\end{widetext}
Regrouping all the terms and using the fact that  $Q(s)=\tilde{\Psi}[\mathbf{k},z=d]$ with $\mathbf{k}=k_0n\sin{\vartheta}\boldsymbol{\hat{\varrho}}$ and $\varrho=r\sin{\vartheta}$ this allow us to obtain a decomposition of the Fourier field $\tilde{\Psi}[\mathbf{k},z=d]$ into a singular (i.e. SP) and regular contribution:
\begin{eqnarray}
\tilde{\Psi}^{\bot,||}[\mathbf{k},z=d]=\tilde{\Psi}^{\bot,||}_{0}[\mathbf{k},z=d]+\tilde{\Psi}^{\bot,||}_{SP}[\mathbf{k},z=d]
\end{eqnarray}
with
\begin{eqnarray}
\tilde{\Psi}^{\bot}_{SP}[\mathbf{k},z=d]=\frac{i\mu_\bot}{8\pi}\frac{k_p}{k_{1,p}}\frac{e^{ik_{1,p}h}e^{ik_{3,p}d}}{\pi\sqrt{kk_p}(k-k_p)}\frac{N_{13}(k_p)}{\frac{\partial D_{13}(k_p)}{\partial k_p} }\nonumber\\
\tilde{\Psi}^{||}_{SP}[\mathbf{k},z=d]=\frac{\boldsymbol{\mu}_{||}\cdot\hat{\mathbf{k}}}{8\pi}\frac{e^{ik_{1,p}h}e^{ik_{3,p}d}}{i\pi\sqrt{kk_p}(k-k_p)}\frac{N_{13}(k_p)}{\frac{\partial D_{13}(k_p)}{\partial k_p} }.
\end{eqnarray}
These formulas are rigorously only valid in the propagative sector where $|\mathbf{k}|\leq k_0n$ (i.e. from the far-field definition). However, due to the simplicity of the  mathematical expressions obtained one is free to extend the validity of Eqs.~65 to the full spectrum of $\mathbf{k}\in \mathbb{R}^2$ values including both  the propagative sector for which $k_3=\sqrt{(k_0^2n^2-|\mathbf{k}|^2)}$ and the evanescent sector for which $k_3=i\sqrt{(|\mathbf{k}|^2-k_0^2n^2)}$ (i.e. if $|\mathbf{k}|\geq k_0 n$).\\ It should now be observed that we can slightly modify our current analysis by observing that Eq.~61 is not exactly a Laurent series  since there are other isolated singularities in the complex plane which were here included in the definition of $Q_0(s)$ i.e. $\tilde{\Psi}^{\bot,||}_{0}[\mathbf{k},z=d]$.  The previous choice was justified for all practical purposes by the detailed calculation done in Section 2 in which only the $\xi_p$ singularity corresponding to the $s_l$ mode contributed to the  integration contours used. Still, for the symmetry of the mathematical expressions it is clearly possible, and actually very useful (as we will see below), to extract a second SP contribution $\xi_{-p}=-\xi_p$ corresponding to $-k_p$. This is clearly the $s_l$ pole associated with propagation in the opposite radial direction. Taking into  account this second pole and the symmetries of $k_{1p}$, $k_{3p}$, $N_{13}(k_p)$ and antisymmetry of $\frac{\partial D_{13}(k_p)}{\partial k_p}$ in the substitution $k_p\rightarrow-k_p$ one obtain after straightforward calculations:
\begin{widetext}\begin{eqnarray}
\tilde{\Psi}^{\bot}_{SP}[\mathbf{k},z=d]:=\frac{i\mu_\bot}{8\pi}\frac{k_p}{k_{1,p}}\frac{e^{ik_{1,p}h}e^{ik_{3,p}d}}{\pi\sqrt{k_p}}\frac{N_{13}(k_p)}{\frac{\partial D_{13}(k_p)}{\partial k_p} }\frac{1}{\sqrt{k}}[\frac{1}{k-k_p}+\frac{1}{i(k+k_p)}]\nonumber\\
\tilde{\Psi}^{||}_{SP}[\mathbf{k},z=d]:=\frac{\boldsymbol{\mu}_{||}\cdot\hat{\mathbf{k}}}{8\pi}\frac{e^{ik_{1,p}h}e^{ik_{3,p}d}}{\pi\sqrt{k_p}}\frac{N_{13}(k_p)}{\frac{\partial D_{13}(k_p)}{\partial k_p} }\frac{1}{\sqrt{k}}[\frac{1}{i(k-k_p)}+\frac{1}{k+k_p}].
\end{eqnarray}\end{widetext}
From this definition we can calculate the SP field in the complete space.   In particular for $z\geq d$ we have $\Psi^{\bot,||}_{SP}(\mathbf{x},z)=\int d^2 \mathbf{k}\tilde{\Psi}^{\bot,||}_{SP}[\mathbf{k},z=d]e^{ik_3Z}$ with $Z=z-d$. More precisely  using the symmetry of the system we obtain
\begin{eqnarray}
\Psi^{\bot}_{SP}(\mathbf{x},z)=\frac{i\mu_\bot}{8\pi}\frac{k_p}{k_{1,p}}\frac{e^{ik_{1,p}h}e^{ik_{3,p}d}}{\pi\sqrt{k_p}}\frac{N_{13}(k_p)}{\frac{\partial D_{13}(k_p)}{\partial k_p} }\nonumber\\
\cdot 2\pi\int_0^{+\infty}\frac{kdk e^{ik_3Z}J_0(k\varrho)}{\sqrt{k}}[\frac{1}{k-k_p}+\frac{1}{i(k+k_p)}]\end{eqnarray} and
\begin{eqnarray}\Psi^{||}_{SP}(\mathbf{x},z)=\frac{\boldsymbol{\mu}_{||}\cdot\hat{\boldsymbol{\varrho}}}{8\pi}\frac{e^{ik_{1,p}h}e^{ik_{3,p}d}}{\pi\sqrt{k_p}}\frac{N_{13}(k_p)}{\frac{\partial D_{13}(k_p)}{\partial k_p} }\nonumber\\
\cdot 2\pi\int_0^{+\infty}\frac{kdk e^{ik_3Z}J_1(k\varrho)}{\sqrt{k}}[\frac{1}{k-k_p}-\frac{1}{i(k+k_p)}].
\end{eqnarray}
with $\mathbf{x}=\varrho\hat{\boldsymbol{\varrho}}$. To obtain these last equations we also used the well known Bessel function properties:
\begin{eqnarray}
\oint d\varphi_ke^{ik\varrho\cos{(\varphi-\varphi_k)}}
\{\begin{array}{c}
\cos{(m\varphi_k)}\\
\sin{(m\varphi_k)}
\end{array}\}\nonumber\\
=2\pi i^m\{\begin{array}{c}\cos{(m\varphi)}\\ \sin{(m\varphi)}\end{array}\}J_m(k\varrho)
\end{eqnarray} (m=0,1,...) to integrate over the $\varphi_k$-coordinate of the 2D vector $\mathbf{k}$.\\ We point out that the convergence of integrals 69, 70 is ensured since the Cosine integral $\int^{+\infty}_{a}dk cos{(k\varrho)}/k=-\textrm{Ci}(a\varrho)\simeq \frac{\cos{(a\varrho)}}{(a\varrho)^2}-\frac{\sin{(a\varrho)}}{a\varrho}$ for $a\varrho\gg 1$ is bounded.
\subsection{Asymptotic expansion}
Remarkably, using the relations $H^{(+)}_0(u)-H^{(+)}_0(-u)=2J_0(u)$ and $H^{(+)}_1(u)+H^{(+)}_1(-u)=2J_1(u)$ (valid for $|\arg{(z)}|< \pi$) as well as the parity properties of the functions $\frac{1}{\sqrt{k}}[\frac{1}{k-k_p}\pm\frac{1}{i(k+k_p)}]$ (i.e. under the transformation $k_p\rightarrow-k_p$) we obtain the practical relations
\begin{eqnarray}
\int_0^{+\infty}\frac{kdk e^{ik_3Z}J_0(k\varrho)}{\sqrt{k}}[\frac{1}{k-k_p}+\frac{1}{i(k+k_p)}]\nonumber\\
=\frac{1}{2}\int_{-\infty}^{+\infty}\frac{kdk e^{ik_3Z}H^{(+)}_0(k\varrho)}{\sqrt{k}}[\frac{1}{k-k_p}+\frac{1}{i(k+k_p)}]\end{eqnarray}
 and
\begin{eqnarray}
\int_0^{+\infty}\frac{kdk e^{ik_3Z}J_1(k\varrho)}{\sqrt{k}}[\frac{1}{k-k_p}-\frac{1}{i(k+k_p)}]\nonumber\\
=\frac{1}{2}\int_{-\infty}^{+\infty}\frac{kdk e^{ik_3Z}H^{(+)}_1(k\varrho)}{\sqrt{k}}[\frac{1}{k-k_p}-\frac{1}{i(k+k_p)}].
\end{eqnarray}
Those relations would not be possible if we didn't included both the
$k_p$ and $-k_p$ poles in the analysis. Inserting Eqs.~72,73 into
Eqs.~69,70 and using the complex variable $\xi$ such as
$k=k_0n\sin{\xi}$ and the integration contour $\Gamma$ used in the
previous Sections we obtain
\begin{eqnarray}
\Psi^{\bot,||}_{SP}(\mathbf{x},z)=\int_{\Gamma}d\xi F^{\bot,||}_{SP}(\xi)e^{ik_0nr\cos{(\xi-\vartheta)}}
\end{eqnarray}
where
\begin{widetext}\begin{eqnarray}
F^{\bot}_{SP}(\xi)=\frac{i\mu_\bot}{8\pi}\frac{k_p}{k_{1,p}}\frac{\sqrt{(k_0n)}e^{ik_{1,p}h}e^{ik_{3,p}d}}{\sqrt{k_p}}\frac{N_{13}(k_p)}{\frac{\partial D_{13}(k_p)}{\partial k_p} }\frac{\sin{\xi}H_0^{(+)}(k_0n\varrho\sin{\xi})e^{-ik_0n\varrho\sin{\xi}}}{\sqrt{\sin{\xi}}}[\frac{\cos{\xi}}{\sin{\xi}-\sin{\xi_p}}+\frac{\cos{\xi}}{i(\sin{\xi}+\sin{\xi_p})}]\nonumber\\
F^{||}_{SP}(\xi)=\frac{\boldsymbol{\mu}_{||}\cdot\hat{\boldsymbol{\varrho}}}{8\pi}\frac{\sqrt{(k_0n)}e^{ik_{1,p}h}e^{ik_{3,p}d}}{\sqrt{k_p}}\frac{N_{13}(k_p)}{\frac{\partial D_{13}(k_p)}{\partial k_p} }\frac{\sin{\xi}H_1^{(+)}(k_0n\varrho\sin{\xi})e^{-ik_0n\varrho\sin{\xi}}}{\sqrt{\sin{\xi}}}[\frac{\cos{\xi}}{\sin{\xi}-\sin{\xi_p}}-\frac{\cos{\xi}}{i(\sin{\xi}+\sin{\xi_p})}].\nonumber\\
\end{eqnarray}\end{widetext}
The integral along $\Gamma$ can be evaluated by using the same
contour deformation as in Section 2. However, due to the absence of
the square root $k_1$ in Eq.~75 there is no branch cut contribution
to the integration contour. The integral can thus be split into one
contribution from the residue and one contribution from the SDP. We
get therefore:
\begin{eqnarray}
\Psi^{\bot,||}_{SP}(\mathbf{x},z)=2\pi i \textrm{Res}[F^{\bot,||}_{SP}(\xi_p)]e^{ik_0nr\cos{(\xi_p-\vartheta)}}\Theta(\vartheta-\vartheta_{LR})\nonumber\\+e^{ik_0nr}\sum_{m\in \textrm{even}}\frac{\Gamma(\frac{m+1}{2})}{m!(k_0nr)^{\frac{m+1}{2}}}\frac{d^m}{d\tau^m}G^{\bot,||}_{SP}(0)\nonumber\\
\end{eqnarray}with $G^{\bot,||}_{SP}(\tau)=F^{\bot,||}_{SP}(\xi)\frac{d\xi}{d\tau}$.\\
Few remarks are here important:\\
(i) First, the singular term $$2\pi i
\textrm{Res}[F^{\bot,||}_{SP}(\xi_p)]e^{ik_0nr\cos{(\xi_p-\vartheta)}}\Theta(\vartheta-\vartheta_{LR})$$
is exactly identical to the pole contribution appearing in Eq.~22.
This results from the equality
$\textrm{Res}[F^{\bot,||}_{SP}(\xi_p)]=\textrm{Res}[F_{+}^{\textrm{TM},\bot,||}(\xi_p)]$
(compare with Eqs.~24-25).\\
(ii) Second, the term $m=0$  in the SDP contribution is dominant in
the far-field regime and leads to
$\Psi^{\bot,||}_{SP}(\mathbf{x},z)=\frac{2\pi
k_0n\cos{\vartheta}}{ir}e^{ik_0
nr}\tilde{\Psi}^{\bot,||}_{SP}[k_0n\sin{\vartheta}\boldsymbol{\hat{\varrho}},z=d]$
as expected.\\
(iii) Third, the decomposition
$Q(s)=Q_0(s)+\textrm{Res}[Q(s_p)]/(s-s_p)+\textrm{Res}[Q(-s_{p}]/(s+s_p)$
leads to \begin{eqnarray}
G^{\bot,||}_{SP}(\tau)=G^{\bot,||}_{SP,0}(\tau)+\frac{\textrm{Res}[G^{\bot,||}_{SP}(\tau_p)]}{\tau-\tau_p}
\nonumber\\+\frac{\textrm{Res}[G^{\bot,||}_{SP}(\tau_{-p}]}{\tau-\tau_{-p}}
\end{eqnarray}
where $\tau_{-p}=-e^{i\pi/4}\sqrt{2}\sin{((\xi_{p}+\vartheta)/2)}$.
Therefore, if we compare with Eqs.~32-38 we see that
$\Psi^{\bot,||}_{SP}(\mathbf{x},z)$ is not exactly equal to
$I_{SP}+I_{SDP}^{\textrm{pole}}$ explicitly defined in Eqs.~37 and
36. More precisely we obtain:
\begin{eqnarray}
\Psi^{\bot,||}_{SP}(\mathbf{x},z)=2i\pi\textrm{Res}[G^{\bot,||}_{SP}(\tau_p)]e^{ik_0nr\cos{(\xi_p-\vartheta)}}\Theta(\vartheta-\vartheta_{LR})\nonumber\\
-e^{ik_0nr}\sum_{m\in \textrm{even}}\frac{\Gamma(\frac{m+1}{2})}{(k_0nr)^{\frac{m+1}{2}}}\frac{\textrm{Res}[G^{\bot,||}_{SP}(\tau_p)]}{\tau_p^{m+1}} \nonumber\\
-e^{ik_0nr}\sum_{m\in \textrm{even}}\frac{\Gamma(\frac{m+1}{2})}{(k_0nr)^{\frac{m+1}{2}}}\frac{\textrm{Res}[G^{\bot,||}_{SP}(\tau_{-p})]}{\tau_{-p}^{m+1}}\nonumber\\
+e^{ik_0nr}\sum_{m\in \textrm{even}}\frac{\Gamma(\frac{m+1}{2})}{m!(k_0nr)^{\frac{m+1}{2}}}\frac{d^m}{d\tau^m}G^{\bot,||}_{SP,0}(0)\nonumber\\
\end{eqnarray} which differs from Eqs.~37, 38 by the two last lines.
We can also rewrite these expressions as
\begin{eqnarray}
\Psi^{\bot,||}_{SP}(\mathbf{x},z)=e^{ik_0nr}\sum_{m\in \textrm{even}}\frac{\Gamma(\frac{m+1}{2})}{m!(k_0nr)^{\frac{m+1}{2}}}\frac{d^m}{d\tau^m}G^{\bot,||}_{SP,0}(0)\nonumber\\
+i\pi\textrm{Res}[G^{\bot,||}_{SP}(\tau_p)]e^{ik_0nr\cos{(\xi_p-\vartheta)}}\textrm{erfc}(-i\tau_p\sqrt{(k_0nr)})\nonumber\\
+i\pi\textrm{Res}[G^{\bot,||}_{SP}(\tau_{-p})]e^{ik_0nr\cos{(\xi_p+\vartheta)}}\textrm{erfc}(-i\tau_{-p}\sqrt{(k_0nr)})\nonumber\\
\end{eqnarray}
where we used Eq.~36. applied to $\tau_{-p}$ and $\tau_{p}$.
\section{More on intensity and field in the back focal plane and image plane of the microscope}
A general analysis of the imaging process occurring through a
microscope objective with high numerical aperture $NA$ and an ocular
tube lens is given in for example Ref.~\cite{Sheppard}.  Here, we
give without proofs the calculated field and intensity in the focal
plane of the
objective and the image plane of the microscope expressed in term of the TE and TM scalar potentials defined in Eqs.~1,2.\\
For this purpose we use the Fourier transform of the electromagnetic
TM and TE field at the $z=d$ interface defined by:
\begin{eqnarray}
\tilde{\mathbf{D}}_{\textrm{TM}}[\mathbf{k},z]
=-\{\mathbf{k}k_3(k)-k^2\hat{\mathbf{z}}\}\tilde{\Psi}_{\textrm{TM}}[\mathbf{k},z]\nonumber\\
\tilde{\mathbf{D}}_{TE}[\mathbf{k},z]
=-k_0n^2\mathbf{k}\times\mathbf{\hat{z}}\tilde{\Psi}_{TE}[\mathbf{k},z].
\end{eqnarray}
This implies \cite{Sheppard} that the electric field recorded in the
back focal plane of the objective is (i.e. taking into account the
vectorial nature of the field and the transformation of the
spherical wave front to a planar wave front):
\begin{widetext}
\begin{eqnarray}
\mathbf{E}_{\textrm{back focal plane $\Pi$}}=\frac{2\pi
e^{ik_0nf}}{if}\frac{T_1\sqrt{k_0k_3(k)}}{n}\{-k_0n\mathbf{k}\tilde{\Psi}_{\textrm{TM}}[\mathbf{k},d]+k_0n^2k\boldsymbol{\hat{\varphi}}_1\tilde{\Psi}_{TE}[\mathbf{k},d]\}
\end{eqnarray}
\end{widetext}with by definition $\boldsymbol{\hat{\varphi}}_1=-\mathbf{k}\times\hat{\mathbf{z}}/k$.
The geometric coefficient $k_3(k)$ is reminiscent from  the 'sin'
condition \cite{Sheppard} which lead to strong geometrical
abberations at very large angle $\theta$.  As a direct consequence
we deduce the intensity in the back focal plane:
\begin{eqnarray}
|\mathbf{E}_{\textrm{back focal plane $\Pi$}}|^2=\frac{4\pi^2t_1}{f^2n^2}k_0k_3(k)[|\mathbf{\tilde{D}}_{\textrm{TM},3}[\mathbf{k},d]|^2+|\mathbf{\tilde{D}}_{\textrm{TE},3}[\mathbf{k},d]|^2]\nonumber\\
\end{eqnarray} which is  therefore proportional to the total Fourier field intensity  for TM and TE waves taken
separately.\\
Finally, in the image
plane we obtain the electric field :
\begin{eqnarray}
\mathbf{E}(\mathbf{x}')=N'\int_{|\mathbf{k}|\leq
k_0NA}d^2\mathbf{k}\sqrt{k_3(k)}e^{-i\mathbf{k}\cdot\frac{\mathbf{x}'}{M}}
\cdot\{\tilde{\mathbf{D}}_{\textrm{TM},||}[\mathbf{k},d]\frac{k_0n}{k_3(k)}+\tilde{\mathbf{D}}_{\textrm{TE}}[\mathbf{k},d]\}\nonumber\\
\end{eqnarray}i.e.
\begin{eqnarray}
\mathbf{E}(\mathbf{x}')=N'\int_{|\mathbf{k}|\leq
k_0NA}d^2\mathbf{k}\sqrt{k_3(k)}e^{-i\mathbf{k}\cdot\frac{\mathbf{x}'}{M}}
\cdot\{-k_0n\mathbf{k}\tilde{\Psi}_{\textrm{TM}}[\mathbf{k},d]+k_0n^2k\boldsymbol{\hat{\varphi}}_1\tilde{\Psi}_{\textrm{TE}}[\mathbf{k},d]\}
\end{eqnarray}
where $N'$ is a constant characterizing the microscope. In the
letter we used theses formulas for computing fields and intensity in
the Fourier and image plane (see Figs.~3,4 of the letter).
\appendix
 \section{}
We have by definition \begin{eqnarray}\tau_p''=\sin{((\xi_p'-\vartheta)/2)}\cosh{(\xi_p''/2)}\nonumber\\
+\cos{((\xi_p'-\vartheta)/2)}\sinh{(\xi_p''/2)}.\end{eqnarray}
The condition $\tau_p''<0$ is equivalent to $\tan{((\xi_p'-\vartheta)/2)}<-\tanh{(\xi_p''/2)}$, i.e. to
\begin{eqnarray}
\frac{\xi_p'-\vartheta}{2}<-\arctan{(\tanh{(\xi_p''/2)})}\nonumber\\=-\arccos{(\frac{1}{\cosh{(\xi_p'')}})}.
\end{eqnarray}
We therefore obtain $\vartheta_{LR}<\vartheta$ where holds the relation \begin{eqnarray}\cos{((\xi_p'-\vartheta_{LR})}\cosh{\xi_p''}=1.\end{eqnarray} This is clearly the definition of the leakage radiation angle introduced in the discussion of the singular term $I_{SP}$. This therefore implies the equality
\begin{eqnarray}\Theta(\vartheta-\vartheta_{LR})=\Theta(-\tau_p'').\end{eqnarray}
 \section{}
 We have the relation $G(\tau)=F(\xi)\frac{d\xi}{d\tau}$ and we define
 \begin{eqnarray}
 F(\xi)=F_0(\xi)+\frac{\textrm{Res}[F(\xi_p)]}{\xi-\xi_p}\nonumber\\
G(\tau)=G_0(\tau)+\frac{\textrm{Res}[G(\tau_p)]}{\tau-\tau_p}
 \end{eqnarray}
 Therefore  we obtain for the residues the relation:
  \begin{eqnarray}
\textrm{Res}[G(\tau_p)]=\frac{1}{2\pi i} \oint_{C_p} d\tau G(\tau)\nonumber\\
=\frac{1}{2\pi i} \oint_{\mathcal{C}_p} d\xi F(\xi)=\textrm{Res}[F(\xi_p)] .
 \end{eqnarray}